\documentclass{IEEEtran}
   
    \usepackage[pdftex]{graphicx}
    \graphicspath{{../pdf/}{../jpeg/}}
	\DeclareGraphicsExtensions{.pdf,.jpeg,.png}

	\usepackage[cmex10]{amsmath}
	\usepackage{mathabx}
	\usepackage{algorithmic}
	\usepackage{array}
	\usepackage{mdwmath}
	\usepackage{mdwtab}
	\usepackage{eqparbox}
    \newcolumntype{P}[1]{>{\centering\arraybackslash}p{#1}}
    \newcolumntype{M}[1]{>{\centering\arraybackslash}m{#1}}
    \usepackage{rotating}
    \usepackage{bm}
 
    \usepackage{tabularx}
    \usepackage{caption}
    \usepackage{mwe}
    \usepackage{graphicx}
    \usepackage{sidecap}
    \usepackage{derivative}
    \usepackage{amsmath}

    \usepackage{setspace}
    \usepackage{xcolor}
    \usepackage{soul}
    \DeclareMathOperator{\taninv}{tan^{-1}}

    \usepackage{cite}
    
\begin{document}

\renewcommand{\thesection}{\arabic{section}}
\renewcommand{\thesubsection}{\arabic{section}.\arabic{subsection}}
\renewcommand{\thesubsubsection}{\arabic{section}.\arabic{subsection}.\arabic{subsubsection}}

\makeatletter
\renewcommand{\@seccntformat}[1]{%
  \csname the#1\endcsname\quad}
\makeatother

\title{\LARGE  A three-axis Nanopositioner based on Near-Field Acoustic Levitation and Electromagnetic Actuation }

\author{ K. S. Vikrant, Prosanto Biswas, and S. O. Reza Moheimani, Fellow, IEEE }

\maketitle

\begin{abstract}
Near-field acoustic levitation (NFAL) enables nanometer-scale positioning resolution and bandwidth exceeding several hundred hertz specifically along the vertical (Z) direction, owing to its high acoustic stiffness and squeeze film damping. However, its application to horizontal (XY) positioning is limited by significantly lower acoustic stiffness and insufficient damping in horizontal directions, resulting in reduced resolution and bandwidth. Moreover, NFAL-based positioning systems typically lack multi-axis actuation capabilities due to challenges in generating multi-directional acoustic forces. This work presents a hybrid positioning approach that overcomes the mentioned limitations by integrating NFAL with electromagnetic actuation. A planar magnetic platform is acoustically levitated, while a coplanar current-carrying coil provides horizontal trapping stiffness more than three orders of magnitude higher than that achievable with acoustic forces alone. Additionally, the coil generates three-dimensional electromagnetic forces, enabling multi-axis positioning capability. Eddy currents induced in a thin copper sheet integrated with the coil enhance horizontal damping by 52 times. We experimentally demonstrate precise 3-axis linear motion with a root mean square (RMS) positioning resolution better than 20~nm along all axes. The system achieves an in-plane motion range of 1.42 mm with a bandwidth of 16 Hz and a Z-axis motion range of 40 µm with a positioning bandwidth of 171 Hz.

\end{abstract}

\IEEEoverridecommandlockouts


\IEEEpeerreviewmaketitle

\section{Introduction}
Precision positioners are critical components in scientific instruments used for imaging, manipulation, fabrication, and characterization of samples \cite{holmes2000long,manske2012recent}. Levitation-based positioning systems are particularly appealing, as they offer high positioning resolution due to the absence of friction and multiple degrees of freedom owing to the lack of mechanical constraints. Among them, self-stabilizing levitation techniques are especially desirable due to their simpler design and compact form factor resulting from self-stability. Several self-stabilizing methods have been explored, including optical \cite{dholakia2011shaping, ilic2019self}, diamagnetic \cite{simon2001diamagnetically,romagnoli2023controlling}, and acoustic levitation \cite{marzo2015holographic, shen2023self,andrade2020acoustic}. Near-Field Acoustic Levitation (NFAL) stands out in particular, as it provides sufficient levitation force to support macroscale platforms, making it highly suitable for the development of multi-axis precision positioning stages.

The NFAL technique has been previously used to develop bearings and motion stages \cite{ide2007non,hashimoto1998transporting, 
 gabai2019contactless,li2024novel}. These systems usually comprise two coplanar platforms, with one stacked on top of the other. The bottom platform is attached to an ultrasonic Z-actuator, while the top platform remains untethered. Due to the high frequency oscillation of the Z-actuator, the air between the the surfaces of the fixed and untethered platforms does not have sufficient time to flow in or out. The time-averaged air pressure within this trapped air, during the cyclic compression and expansion, exceeds the atmospheric pressure because of the nonlinear dependence of the air pressure on the volume of trapped air. This increased air pressure levitates the untethered planar platform a few hundred microns above the bottom platform. Furthermore, compressed air provides significant squeeze film damping along the Z-axis \cite{ilssar2015slow,wang2021stiffness,li2022study}.

Despite its advantages, such as high payload capacity, high trapping stiffness, and damping along the Z-axis, the NFAL technique faces three significant limitations that must be addressed before it can be used to develop multi-axis precision positioners. First, conventional NFAL technique generates ultra-low in-plane trapping stiffness, usually a few millinewton per meter, due to the small magnitude of acoustic restoring forces in the XY-plane\cite{aono2019increase,aono2022measurement,li2024contact}. Second, the NFAL technique provides low in-plane damping due to limited air resistance in the XY-plane. The combination of low stiffness and damping, along with inherent acoustic noise in NFAL systems, degrades the XY positioning resolution to a few millimeters \cite{gabai2019contactless,kikuchi2021development}. Third, the NFAL technique generates predominantly Z-forces, providing only vertical positioning capabilities.  

To address the first two limitations, several geometric designs have been proposed that leverage viscous acoustic forces to enhance trapping stiffness and damping in the XY-plane \cite{aono2019increase,li2024contact}. One such design is the sandwich configuration, in which the levitating platform is positioned between two oscillating platforms. This configuration increases the in-plane trapping stiffness from approximately 10\,mN/m to 20\,mN/m near the stable equilibrium position \cite{aono2019increase}. More recently, a negative pressure adsorption effect has been introduced, achieving a nearly four-fold increase in trapping stiffness from about 10\,mN/m to 40\,mN/m \cite{li2024contact}. Both approaches also yield a two- to three-fold improvement in horizontal damping. While these enhancements in trapping stiffness and damping make NFAL suitable for coarse positioning tasks with millimeter-scale accuracy, they remain inadequate for high-precision micro- and nano-positioning applications.

To address the third limitation, namely, the lack of multi-axis actuation capabilities, researchers have explored the use of acoustic traveling waves generated by flexural vibrations of the bottom platform \cite{hashimoto1998transporting,gabai2019contactless}. These traveling waves exert in-plane forces on the levitating platform, generating XY motion. Since the platform remains in neutral equilibrium in the XY-plane, a multi-axis measurement and feedback control system is required to stabilize the motion, thereby enabling two-axis precision positioning over centimeter-scale ranges \cite{gabai2019contactless}. More recently, an innovative positioning approach has been reported based on an array of oscillating bottom platforms and a single levitating platform. Here, the levitating platform hops from one bottom platform to another as the bottom platforms are sequentially oscillated, resulting in XY-motion \cite{kikuchi2021development}. Although these methods provide in-plane positioning capabilities, their positioning resolution and bandwidth remain limited, typically at the millimeter scale and sub-hertz range, due to the small magnitude, inherent fluctuations, and slow dynamics of the in-plane acoustic forces used to drive the levitating platform \cite{gabai2019contactless,li2024novel,kikuchi2021development}.

This paper addresses the three aforementioned limitations by introducing a hybrid positioning system that combines Near-Field Acoustic Levitation (NFAL) with electromagnetic stabilization. Specifically, an ultrasonic Z-actuator levitates an untethered planar magnetic platform using near-field acoustic forces, while a co-planar current-carrying coil positioned beneath the levitating platform addresses two of the limitations. First, the coil generates steady electromagnetic forces that enhance horizontal trapping stiffness by over three orders of magnitude compared to prior NFAL systems. The resulting stiffness of approximately \( 40\, \text{N/m} \) enables nanometer-scale horizontal positioning resolution, representing an improvement of more than four orders of magnitude compared to the sub-millimeter-scale XY-resolution achieved by previously reported NFAL positioners. Additionally, the high electromagnetic trapping stiffness in the XY-plane allows the hybrid positioner to achieve two orders of magnitude higher in-plane positioning bandwidth \cite{gabai2019contactless,li2024novel,kikuchi2021development}. Second, the same planar coil applies three-dimensional (3D) electromagnetic forces, enabling multi-axis actuation without requiring feedback stabilization. Finally, to address the third limitation, i.e., the limited air resistance and insufficient damping during horizontal motion, we introduce an eddy current-based damping mechanism. This is implemented by integrating a thin, co-planar copper sheet above the stationary planar coil. The resulting eddy currents lead to a 52-fold increase in in-plane damping. Experimentally, we demonstrate in-plane motion over a range of $1.42\,\text{mm}$ and out-of-plane motion of $40\,\mu\text{m}$, with positioning precision better than $20\,\text{nm}$ along all three axes. The actuation bandwidth achieved is $16\,\text{Hz}$ in-plane and $171\,\text{Hz}$ along the Z-axis.

The structure of the remainder of this paper is as follows: Section 2 presents an overview of the positioner. Section 3 describes the modeling of electromagnetic and acoustic forces. Section 4 discusses the evaluation of the developed positioner, and finally, Section 5 concludes the paper.

\section{Overview}

\begin{figure}
    \begin{centering}
        \includegraphics[width=0.41\paperwidth]{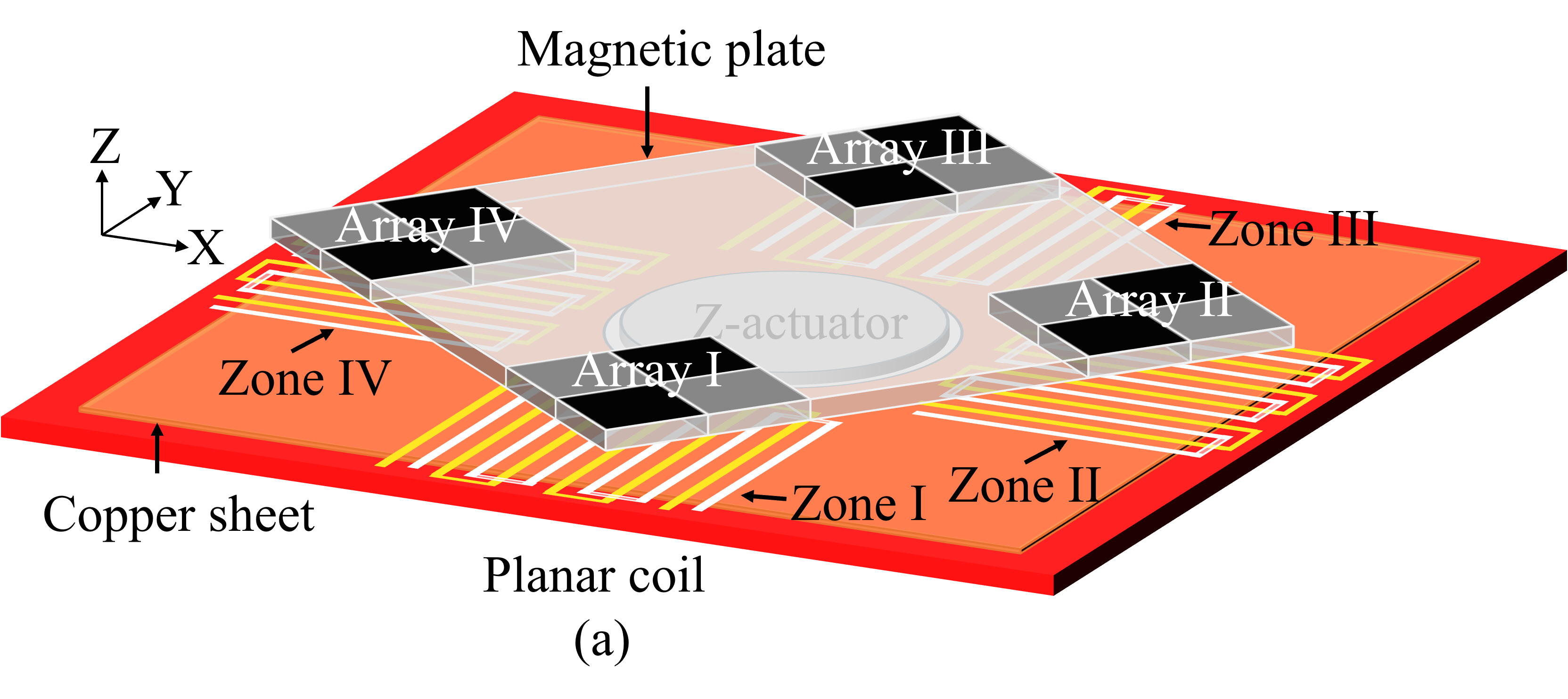}
        \includegraphics[width=0.41\paperwidth]
        {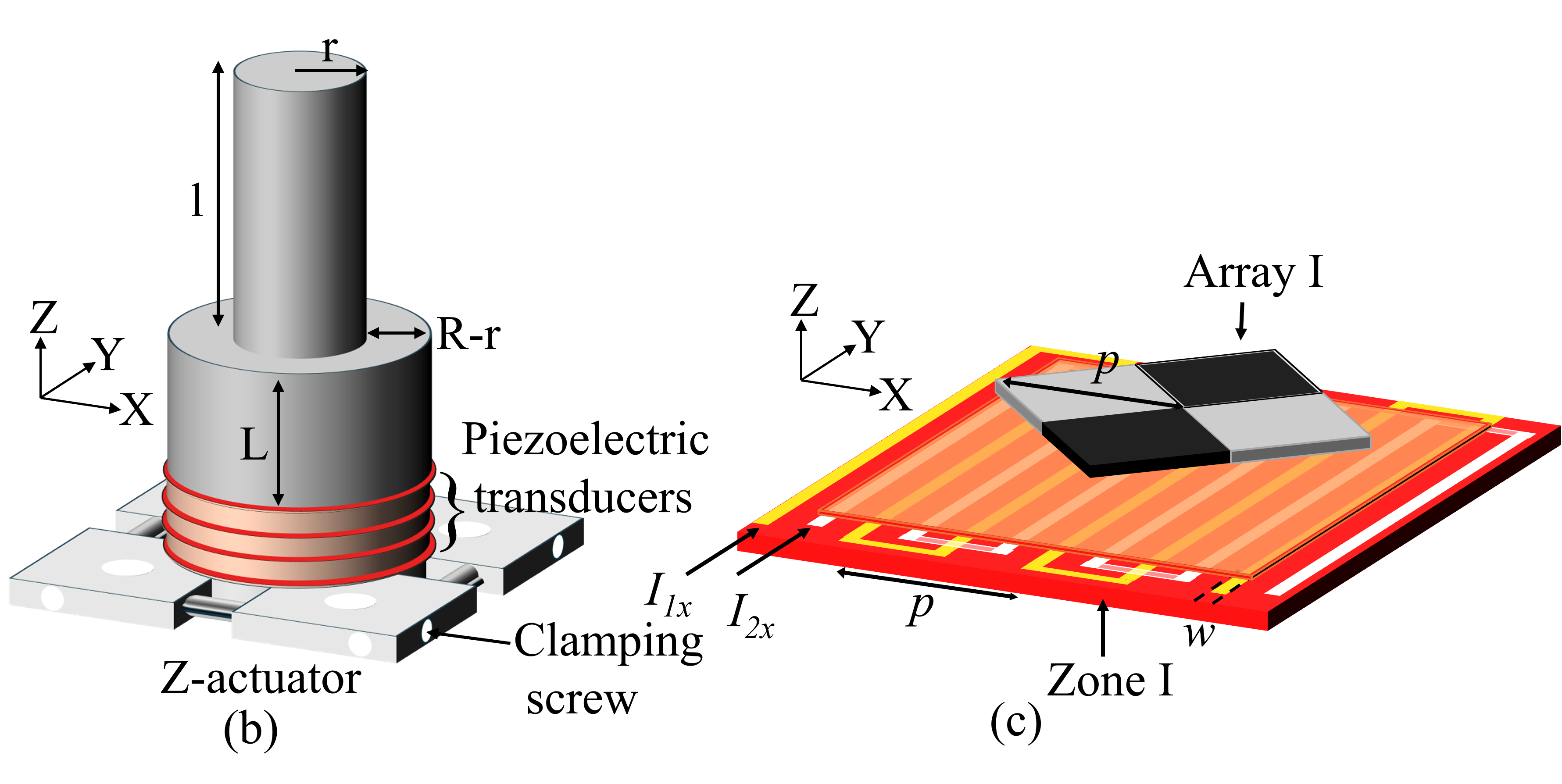}
    \par
    \end{centering}
    \caption{\small Schematic showing (a) the positioner, which consists of the Z-actuator, the magnetic plate, the planar current-carrying coil, and the copper sheet; (b) the Z-actuator, comprising a mechanical horn clamped to piezoelectric transducers; and (c) the current-carrying tracks in Zone I of the planar coil, with the magnetic array I positioned directly above Zone I.}
    \label{Overview}
    \vspace{-10pt} 
\end{figure}

Fig.~\ref{Overview}(a) shows a schematic of the proposed positioner, which comprises a Z-actuator, a magnetic plate, a planar current-carrying coil, and a co-planar copper sheet integrated above the coil. The Z-actuator levitates the magnetic plate, providing high acoustic stiffness and squeeze film damping in the vertical (Z) direction. The planar coil generates electromagnetic forces that trap the levitated plate, offering large in-plane restoring forces and high trapping stiffness in the XY-plane. Additionally, the coil produces 3D electromagnetic forces, enabling multi-axis positioning capabilities. Finally, the integrated copper sheet induces eddy currents that significantly damp horizontal (XY) motion of the levitating magnetic plate.

The Z-actuator consists of three annular piezoelectric transducers of radius $R$, bonded to a stepped mechanical horn [Fig.~\ref{Overview}(b)]. The horn is formed by two concentric cylinders with radii $R$ and $r$, and heights $L$ and $l$, respectively. The end of the horn with radius $R$ is clamped to the piezoelectric elements, while the end with radius $r$ is free. When driven at its ultrasonic resonance frequency, the horn amplifies the oscillation amplitude, producing a vibration amplitude significantly larger than that achievable directly from the piezoelectric transducers alone. This ultrasonic Z-motion traps air in the overlapping region of area $\pi r^2$  between the actuator and the magnetic plate. Periodic compression and expansion of the trapped air increase the average air pressure above atmospheric levels due to the nonlinear relationship between pressure and volume of the trapped air. At sufficiently high oscillation amplitudes, the resulting acoustic pressure generates enough force to levitate the magnetic plate. The trapped air also generates high acoustic stiffness and squeeze film damping, providing high-bandwidth vertical positioning capability.

The levitating plate is made from non-ferrous material embedded with four checkerboard pattern magnetic arrays I, II, III, and IV. Each checkerboard array is designed using $n$ square permanent magnets, where each permanent magnet has diagonal length $p$, thickness $t$, and magnetic moment $m$ aligned along the Z-axis. The black and gray colors in the checkerboard pattern represent magnets with opposite magnetic moments.  In Fig.~\ref{Overview}, $n$ is selected as four for illustrative purposes. The actual number $n$ depends on the dimensions of both the magnet and the plate. 

The planar electromagnetic coil comprises four Zones I, II, III, and IV, where Zone I is opposite Zone III and Zone II is opposite Zone IV. Furthermore, the current-carrying tracks in Zones I and III are orthogonal to the current-carrying tracks in Zones II and IV. Each zone stabilizes and actuates the magnetic array positioned directly above that zone. The opposite Zones I and III electromagnetically trap the Arrays I and III, respectively, along the X-axis. In contrast, the opposite Zones II and IV electromagnetically trap the Arrays II and IV, respectively, along the Y-axis. Thus, together, these four zone-array pairs stabilize the levitating plate in the XY-plane. 

Since the four Zone-Array pairs are identical, one of the pairs, namely Zone I-Array I, is shown in detail in Fig.~\ref{Overview}(c). Zone I consists of two current-carrying tracks, \( I_{1x} \) and \( I_{2x} \). Track \( I_{1x} \) consists of multiple thin, long conductors of width \( w \), aligned parallel to the Y-axis, and connected by short conductors along the X-axis at their extremities. The conductors of the track \( I_{1x} \) that carry current in the same direction are separated by a distance \( p \), the pitch of the coil. Track \( I_{2x} \) is identical to the track \( I_{1x} \), with the only difference being an X-offset of \( p/4 \). The checkerboard pattern of Array I is designed so that magnets with identical magnetic moments are separated by a distance \( p \), the pitch of the magnetic array. This equal pitch for both the coil and the array ensures that all magnets in Array I experience identical electromagnetic forces, as derived in the modeling section discussed later.

To achieve linear X-motion, opposite Zones I and III apply identical X-forces on Arrays I and III, respectively. Similarly, opposite Zones II and IV apply identical Y-forces on Arrays II and IV, respectively, to actuate the levitating plate along the Y-axis. The levitating plate can be translated along the Z-axis using either electromagnetic or acoustic Z-forces.  The electromagnetic Z-forces for linear Z-motion are generated using all four zones or any pair of opposite zones. Next, to generate in-plane rotational motion, the opposite zones apply equal but opposite in-plane forces to produce a couple, thereby rotating the levitating plate about the Z-axis. Similarly, to generate out-of-plane rotational motion, the opposite zones apply equal but opposite out-of-plane forces to produce a couple, thus rotating the levitating plate about the X- and Y-axes. Therefore, the four zone-array pairs provide six-axis positioning capability. In this paper, we only report 3-axis linear positioning capability. The angular positioning capability will be explored in future research work.

\section{Modeling}
Section 3.1 describes the analytical model developed for calculating 3D actuation forces and trapping stiffness. The dynamic model of the positioner is discussed in Section 3.2.  
\vspace{-1 em}
\subsection{Actuation forces and trapping stiffness}

\begin{figure}[hbt!]
    \begin{centering}
        \includegraphics[width=0.42\paperwidth]{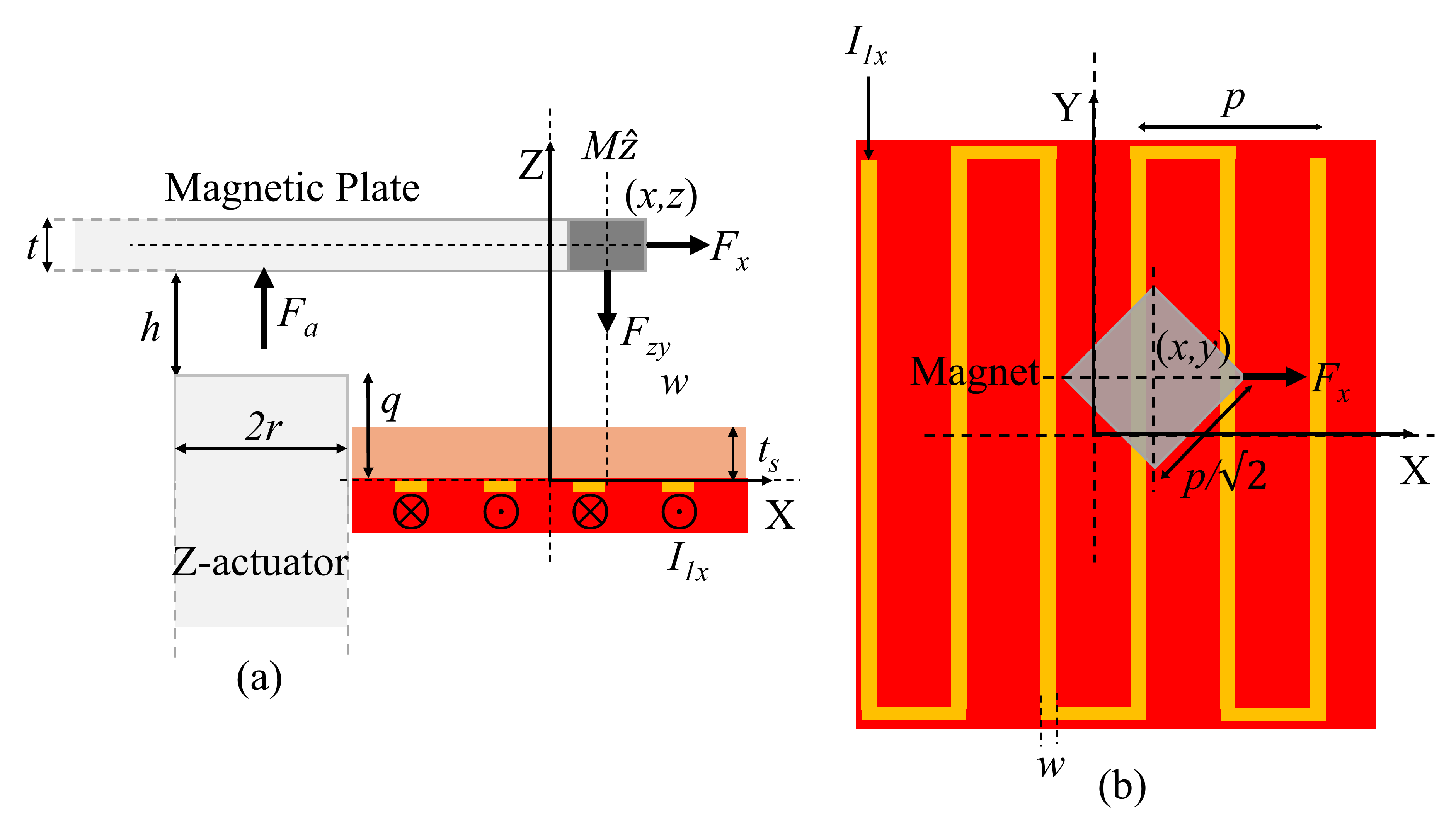}
    \par
    \end{centering}
    \caption{  \small Schematic showing (a) the side view of the positioning system. For acoustic modeling purposes, only a portion of the magnetic plate and the Z-actuator are shown, along with a single magnet attached to the magnetic plate and one of the track currents. (b) The top view showing one of the tracks in Zone I and a single magnet from Array I, represented for electromagnetic modeling purposes.} 
    \label{modeling}
\end{figure}

The schematic shown in Fig.~\ref{modeling} is used to calculate the actuation forces and trapping stiffness along all three axes. In the XY-plane, both the actuation forces and stiffness arise primarily from electromagnetic interactions between the planar coil and the magnetic plate. Due to the negligible magnitude of in-plane acoustic forces, they are not included in the model. In contrast, the actuation force and stiffness along the Z-axis result from the combined effects of electromagnetic and acoustic interactions. Accordingly, we first obtain the electromagnetic force and stiffness components along all three axes. We then calculate the acoustic contributions along the Z-axis and add them to the corresponding electromagnetic components to obtain the total Z-force and Z-stiffness.

The magnetic field generated by the planar current-carrying coil is derived to calculate the electromagnetic forces acting on the magnetic plate. Since the positioner comprises four similar current zone-magnetic array pairs, it suffices to calculate the field and force for one zone-array pair. Here we consider Zone I-Array I for electromagnetic modeling. Furthermore, since both tracks in Zone I are identical, differing only by an X-offset of $p/4$, one of the tracks $I_{1x}$ is shown for modeling. Similarly, since the magnets in array I are identical, only one of the magnets is shown for modeling. The electromagnetic force $\bm{F}$ on the single magnet whose center is positioned at $x,y,z$ is calculated first. Then the forces on other magnets in an array will be calculated by substituting their respective positions in the obtained force expression $\bm{F}$. Finally, these forces are added together to obtain the resultant force acting on the magnetic plate.

Each magnet is made of a material with uniform magnetization $M\hat{z}$.  To calculate the force $\bm{F}$, the selected magnet of volume $V=p^2t/2$ is divided into multiple infinitesimal elements of volume $dV$, each with magnetic moment $\bm{m}= MdV\hat{z}$. These elements experience a force $\bm{dF}$ due to the magnetic field  $\bm{B} = [ B_x\, B_y\, B_z]^T$, generated by the current-carrying tracks in Zone I, expressed as,
\begin{equation} \label{dF}
\bm{dF} = \nabla(\bm{m.B})=\nabla(mB_{z})
\end{equation}
The total electromagnetic force $\bm{F} = [ F_x\, F_y\, F_{zy}]^T$, is obtained by integrating the differential force over the volume of the magnet,
\begin{equation} \label{F}
\bm{F}= \int_V M\nabla B_zdV.
\end{equation}
 To calculate the force using Equation \ref{F}, we require the magnetic field $B_z = B_{z1}I_{1x} + B_{z2}I_{2x}$ where $B_{z1}$ and $B_{z2}$ represents the Z-magnetic field per unit current due to the tracks $I_{1x}$ and $I_{2x}$ in Zone I. Note that the second track is identical to the first but offset by a distance $p/4$. Assuming both tracks comprise a large number of infinitely long conductors parallel to the Y-axis, the field $B_z$ in a region away from the boundaries of Zone 1 and $z\geq p/4$ is found to be 
\begin{equation} \label{Bz}
B_{z} = \frac{\mu_0}{p} \, \text{sinc}(\pi \tilde{w}) \, \text{sech}(2\pi \tilde{z}) \left[ \cos(2\pi \tilde{x}) I_{1x} + \sin(2\pi \tilde{x}) I_{2x} \right]
\end{equation}
where $\tilde{w}$ = $w/p$, $\tilde{z}$ = $z/p$ and $\tilde{x}$ = $x/p$ \cite{10589479}. On substituting the $B_z$ in Equation \ref{F}, the electromagnetic force, $\bm{F} = [F_{x}\quad 0\quad F_{zy}]^T$, and the electromagnetic stiffness, $\bm{k}= [k_{x}\quad 0\quad k_{zy}]^T$, due to the tracks $I_{1x}$ and $I_{2x}$ are found to be \cite{10589479}

\begin{equation} \label{Fvect}
\frac{-\pi^2}{2\mu_0Mp\text{sinc}(\pi\tilde{w})}\bm{F} = \bm{CI_x},
\end{equation}

\begin{equation} \label{kvect}
\frac{\pi}{4\mu_0M\text{sinc}(\pi\tilde{w})}\bm{k} = \bm{DI_x},
\end{equation}

where
\begin{align*}
\bm{C} &= \begin{bmatrix}
\phi \sin(2\pi \tilde{x}) & -\phi \cos(2\pi \tilde{x}) \\
0 & 0 \\
\psi \cos(2\pi \tilde{x}) & \psi \sin(2\pi \tilde{x})
\end{bmatrix}, \quad
\bm{I_x} = \begin{bmatrix} I_{1x} \\ I_{2x} \end{bmatrix}, \\
\bm{D} &= \begin{bmatrix}
\phi \cos(2\pi \tilde{x}) & \phi \sin(2\pi \tilde{x}) \\
0 & 0 \\
\chi \cos(2\pi \tilde{x}) & \chi \sin(2\pi \tilde{x})
\end{bmatrix}, \quad
k_i = -\frac{\partial F_i}{\partial i}(i=x,zy),
\end{align*}
\noindent
$\phi = \tan^{-1}[\sinh(2\pi(\tilde{z}+\tilde{t}/2))]-\tan^{-1}[\sinh(2\pi(\tilde{z}-\tilde{t}/2))],$
$\psi = \text{sech}(2\pi(\tilde{z}-\tilde{t}/2))-\text{sech}(2\pi(\tilde{z}+\tilde{t}/2))$, and 
$\chi = \text{sech}(2\pi(\tilde z +\tilde t/2))\text{tanh}(2\pi(\tilde z +\tilde t/2))-\text{sech}(2\pi(\tilde z -\tilde t/2))\text{tanh}(2\pi(\tilde z -\tilde t/2))$.

The force $\bm{F}$ acting on the selected magnet is a periodic function of its position $x$ with period $p$ as $\bm{C}$ is a periodic function of $x$ with fundamental period $p$. Since the other magnets with identical magnetic moments in Array I are separated by a distance of $p$ along the X-axis, each of them will experience the same force given by Equation \ref{Fvect}. Therefore, the resultant electromagnetic force and stiffness for Array I, which comprises $n$ magnets, are $n\bm{F}$ and $n\bm{k}$, respectively. Additionally, Equation \ref{Fvect} indicates that Zone I applies forces along both X- and Z- axes on Array I. Because Zone III is identical to Zone I, it will exert the same X- and Z- forces on Array III, which can be obtained using Equation \ref{Fvect}, where $I_{1x}$ and $I_{2x}$ will be substituted by the currents through the first and second tracks of Zone III, respectively. 

However, Zones II and IV are rotated by $90^\degree$ in the XY-plane compared to Zones I and III. Consequently, these zones will exert forces along the Y- and Z-axes on Arrays II-IV. Furthermore, because all pair of Zone-Arrays are identical, the Y-force, $F_y$ and the Z-force $F_{zx}$, on each of the magnets in Arrays II-IV  are obtained by substituting $x$ with $y$ in Equation \ref{Fvect} and replacing the current vector $\bm{I_x}$ with the current vector $\bm{I_y} =  [I_{1y}\quad I_{2y}]^T$. Assuming identical currents through opposite zones, the total electromagnetic forces along the X-, Y-, and Z- axes are $2nF_x$, $2nF_y$, and $2nF_{zy}+ 2nF_{zx}$, respectively. It is worth noting that the electromagnetic X- and Y- forces on the magnets are applied exclusively by the tracks parallel to the Y- and X- axes, respectively. However, the electromagnetic Z-force on the magnets is applied by tracks parallel to both the X- and Y- axes. Therefore, to distinguish between these two Z-forces, the Z-forces due to the tracks parallel to the Y- and X- axes are represented by $F_{zy}$ and $F_{zx}$, respectively. Similarly, the total electromagnetic stiffness along the X-, Y- and Z- axes is $2nk_x$, $2nk_y$ and $2nk_{zy}+2nk_{zx}$, respectively. Here, $k_y$ and $k_{zx}$ are obtained by substituting $x$ with $y$ in Equation \ref{kvect} and replacing the current vector $\bm{I_x}$ with the current vector $\bm{I_y} =  [I_{1y}\quad I_{2y}]^T$. After deriving the electromagnetic forces and stiffnesses, we now calculate the acoustic force and stiffness required for levitation and vertical (Z-axis) stability.

The magnetic plate experiences an upward acoustic force \( F_a \) resulting from the high-frequency oscillation of the Z-actuator. This oscillation traps air within the overlapping area \( \pi r^2 \) between the magnetic plate and the actuator, where \( r \) denotes the radius of the actuator's top surface. The pressure distribution within this trapped air varies with both time and radial position along the actuator’s bottom surface. This spatiotemporal pressure distribution is typically computed by solving the compressible Reynolds equation \cite{wang2021stiffness}. Solving this equation yields the time-dependent pressure distribution, from which the instantaneous acoustic force on the magnetic plate is obtained by integrating the pressure over the overlapping surface. Finally, the acoustic Z-force \( F_a \) is computed by temporally averaging this instantaneous force over one oscillation cycle. Although the Reynolds equation can be numerically solved for the configuration shown in Fig.~\ref{modeling}, approximate closed-form expressions for the acoustic force and stiffness derived previously for a similar configuration can be directly applied here.

These expressions for the acoustic Z-force and Z-stiffness were originally developed for an aluminum disk levitating above an oscillating coplanar disk \cite{wang2021stiffness}. Applying those results to the magnetic plate configuration shown in Fig.~\ref{modeling}, the acoustic Z-force and stiffness are given by \( F_a = \frac{\pi r^2 P_a \xi a^2}{2h^2} \) and \( k_a = \frac{\pi r^2 P_a \xi a^2}{h^3} \), respectively. Here, \( a \) is the amplitude of the high-frequency oscillation of the Z-actuator, \( P_a \) is the atmospheric pressure, and \( h \) is the mean separation between the magnetic plate and the Z-actuator surfaces. The parameter \( \xi \) is a dimensionless coefficient that depends on the actuation frequency \( f_d \); its value is approximately unity for ultrasonic frequencies.
To express the acoustic force as a function of the Z-position of the magnetic plate, the separation height is defined as \( h = z - t/2 - q \), where \( q \) denotes the vertical distance between the planar coil and the top surface of the stationary Z-actuator (see Fig.~\ref{modeling}).

\vspace{-6pt} 
\subsection{Dynamic Modeling of the positioner}

First, we determine the stable equilibrium position of the levitating plate ($x_o\,,y_o\,,z_o$), where the resulting forces are zero and the total stiffness is positive in all directions. Along the X-axis, the net force is $2nF_x$ and the total stiffness is $2nk_x$. The force $F_x$ and stiffness $k_x$, obtained from Equations (\ref{Fvect}-\ref{kvect}), are given by

\begin{equation}
\begin{bmatrix}
  F_x\\
  k_x\\  
\end{bmatrix} =  \left( \frac{2\mu_0 M }{\pi^2} \right) I_x p\phi \, \text{sinc}(\pi \tilde{w}) \begin{bmatrix}
  -\sin\left( \frac{2\pi}{p}(x - x_o) \right)\\
  \frac{2\pi}{p} \cos\left( \frac{2\pi}{p}(x - x_o) \right)\\  
\end{bmatrix}
\label{Fxkx}
\end{equation}

where, $I_x = \sqrt{I_{1x}^2 + I_{2x}^2}$ and $x_o = (p/2\pi)\taninv(I_{2x}/I_{1x})$. Using Equation (\ref{Fxkx}), we find that for $x \in [(2u-1)p/2, (2u+1)p/2]$, where $u$ is an integer, position $x = x_o + up$ represents stable equilibrium position, as the net force $2nF_x = 0$ and the total stiffness $2nk_x = k_{xo} = (8n\mu_0 M I_x/\pi) \phi \text{sinc}(\pi\tilde{w}) > 0$. Similarly, the stable equilibrium position along the Y-axis for $y \in [(2u -1)p/2, (2u+1)p/2]$ is determined to be $y = y_o + up$, where $y_o = (p/2\pi)\taninv(I_{2y}/I_{1y})$. The electromagnetic Y-stiffness at the stable equilibrium position is $2nk_y = k_{yo} = (8n\mu_0 M I_y/\pi)\phi \text{sinc}(\pi\tilde{w})$, where $I_y = \sqrt{I_{1y}^2 + I_{2y}^2}.$

Only electromagnetic forces are considered when determining the in-plane stable equilibrium position \( (x_o, y_o) \), as no other forces act on the levitating plate in the XY-plane. However, to determine the out-of-plane stable equilibrium position \( z_o \), we must also account for the acoustic force \( F_a \) and the gravitational force \( m_p g \), in addition to the electromagnetic Z-force acting on the plate. Here, \( m_p \) denotes the mass of the magnetic plate. Therefore, the net Z-force, given by \( F_z = F_a + 2n(F_{zy} + F_{zx}) - m_p g \), must be equal to zero, and the total Z-stiffness, given by \( k_z = k_a - (k_x + k_y) \), must be positive at \( z = z_o \) (see Supplementary Note 1). Since no closed-form solution exists for \( z_o \), it is determined numerically. The total Z-stiffness at \( z = z_o \) is then given by \( k_{zo} = k_a(h_o) - \left(k_{xo}(z_o) + k_{yo}(z_o)\right) \), where \( h_o = z_o - q - t/2 \) represents the mean levitation height of the magnetic plate above the Z-actuator (see Fig.~\ref{modeling}a). Assuming uniform trapping stiffness along the X-, Y-, and Z-axes, the maximum values are calculated as \( k_{xo} = k_{yo} = k_{zo} = k_a / 3 \).

Next, inertial and damping effects are included to obtain the dynamic model of the levitating plate near the stable equilibrium position $(x_o,y_o,z_o)$. For small displacements from equilibrium, the dynamic behavior resembles that of a mass-spring-damper system, where in-plane and out-of-plane damping arises due to the eddy currents and squeeze film, respectively. If $b_e$ and $b_s$ represent the damping coefficients due to the eddy currents and squeeze film, the dynamic model along the X-, Y- and Z- axes can be expressed as:
\begin{equation} \label{xdynamic}
m_{p}\ddot{x} + b_e\dot{x}+ k_{xo}(x-x_o) = F_{xact},
\end{equation}
\begin{equation} \label{ydynamic}
m_{p}\ddot{y} + b_e\dot{y}+ k_{yo}(y-y_o) = F_{yact},
\end{equation}
\begin{equation} \label{zdynamic}
m_{p}\ddot{z} + b_s\dot{z}+ k_{zo}(z-z_o) = F_{zact}.
\end{equation}
Here, $F_{xact} \approx \frac{p k_{xo}I_{1x}}{2\pi I_x^2}\Delta I_{2x}$, $F_{yact} \approx \frac{p k_{yo}I_{1y}}{2\pi I_y^2}\Delta I_{2y}$, and $F_{zact} \approx \frac{\pi r^2P_a\xi a}{h_o^2}\Delta a -\frac{4n\mu_0 Mp}{\pi^2}\text{sinc}(\pi \tilde\omega)\psi(z_o)$
$[\Delta I_{x} + \Delta I_{y}] $ represent small actuation forces along the X-, Y-, and Z- axes, respectively. Therefore, small in-plane displacements with constant trapping stiffness can be achieved by applying a slight variation in actuation currents \((\Delta I_{2x}, \Delta I_{2y})\). Similarly, minute Z-displacement with constant trapping stiffness can be obtained by varying either the acoustic or electromagnetic forces by small amounts, using minute changes in oscillation amplitude \(\Delta a\) and actuation currents\((\Delta I_{x}, \Delta I_{y})\), respectively. Since the 3D actuation forces depend on different actuation parameters, they provide independent positioning capabilities without cross-axis motion.
  
To provide large in-plane displacement, the stable equilibrium position $x_o = (p/2\pi)\taninv(I_{2x}/I_{1x})$ and $y_o = (p/2\pi)\taninv(I_{2y}/I_{1y})$ can be varied with a constant velocity $v_{xo} = p\omega_x/2\pi,v_{yo} = p\omega_y/2\pi$ using sinusoidal actuation currents $I_{1x} = I_{xo}\text{cos}(\omega_x t), I_{2x} = I_{xo}\text{sin}(\omega_x t)$, $I_{1y} = I_{yo}\text{cos}(\omega_y t)$ and $I_{2y} = I_{yo}\text{sin}(\omega_y t)$.  The stiffness of the electromagnetic trap remains constant at $k_{xo}= (8n\mu_0 M I_{xo}/\pi) \phi \text{sinc}(\pi\tilde{w})$ and $k_{yo} = (8n\mu_0 M I_{yo}/\pi) \phi \text{sinc}(\pi\tilde{w})$ during the quasi-static motion. The Z-position of the levitating platform does not vary during the in-plane motion as the net electromagnetic Z-force remains constant at $-\frac{4n\mu_0 Mp}{\pi^2}\text{sinc}(\pi \tilde\omega)\psi(z_o)(I_{xo}+I_{yo})$. Similarly, to provide a large vertical displacement, the Z stable equilibrium position, $z_o$, can be changed by varying either the oscillation amplitude $a$ or the actuation currents $I_x$ and $I_y$. Since $x_o = (p/2\pi)\taninv(I_{2x}/I_{1x})$ and $y_o = (p/2\pi)\taninv(I_{2y}/I_{1y})$, the in-plane position of the levitating plate will remain fixed provided currents through the first and second tracks are changed in the same proportion to achieve the large range Z-motion. However, the 3D trapping stiffness will vary because of the change in both acoustic and electromagnetic stiffness during the large range Z-motion.  

\section{Development and Evaluation}

\begin{figure}[hbt!]
    \begin{centering}
       \includegraphics[width=0.41\paperwidth]{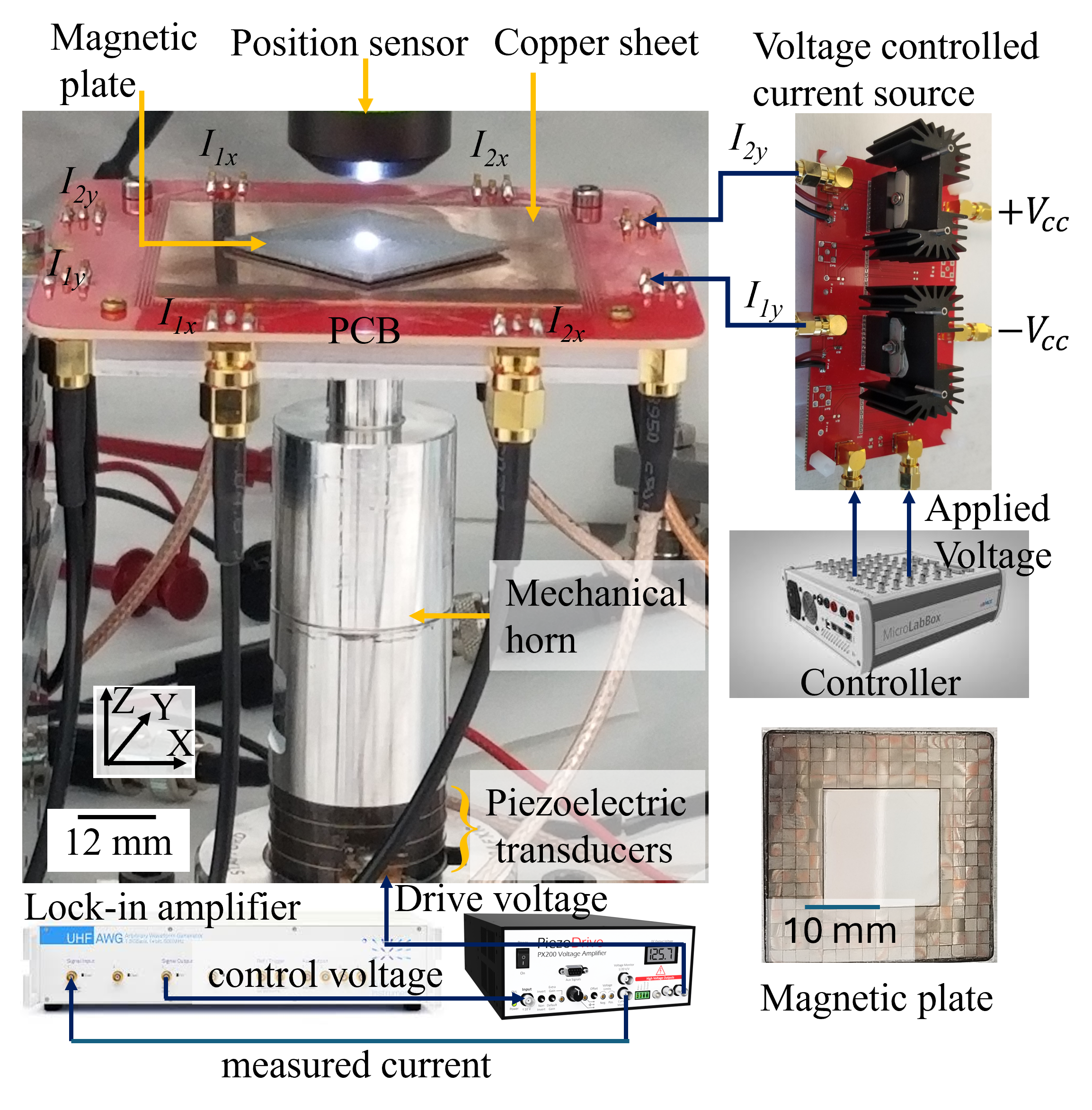}
    \par
    \end{centering}
    \caption{\small Photograph of the positioner showing the Z-actuator, which is designed from a mechanical horn and piezoelectric transducers, the planar coil developed on a PCB with an integrated copper sheet, the magnetic plate, the controller, and the drive electronics. The bottom right of the image displays the rear side of the magnetic plate, including the central polished surface and the checkerboard magnetic array.}
    \label{Positioner}
\end{figure}

The positioner, comprising the magnetic plate, Z-actuator, planar current-carrying coil, and copper sheet, was fabricated and assembled in the configuration discussed in the overview section (Fig.~\ref{Positioner}). The square plate, with an edge length of $30.6\, \text{mm}$ and a thickness of $1.5\, \text{mm}$, was fabricated from aluminum 7075. It includes four rectangular slots along the periphery of its bottom surface to accommodate checkerboard-patterned magnetic arrays, which collectively consist of 208 magnets. Thus, each of the four quadrants of the magnetic plate contains $n = 52$ magnets. The magnets are made from a Neodymium-Iron-Boron (NdFeB) alloy with a pitch length of $p = 2.54\, \text{mm}$, thickness of $t = 0.5\, \text{mm}$, and magnetization of $M =1.27 \times 10^6\, \text{A/m}$ (grade N52). The central portion of the bottom surface of the aluminum plate is polished to an average roughness of $10\, \text{nm}$ to generate sufficient near-field acoustic forces required for levitation. The mass of the aluminum plate with attached magnets is  $m_p = 5.4\, \text{grams}$.
 
The mechanical horn of the Z-actuator was designed from Aluminum 7075 with radii $R = 15\, \text{mm}$, $r = 6\, \text{mm}$, and lengths $L = 60.90\, \text{mm}$ and $l = 28\, \text{mm}$. The upper surface of the Z-actuator was also polished to an average roughness of $10\, \text{nm}$ to generate sufficient upward acoustic forces. A Langevin-type ultrasonic transducer (HEC-3039P4B, Honda Electronics), consisting of three annular piezoelectric transducers, driven by a high-voltage amplifier (PX200-P150, PiezoDrive), is used to actuate the mechanical horn at its first Z resonance frequency of $39.59\, \text{kHz}$. The horn amplifies the Z-motion of the piezoelectric transducers by a factor of approximately $100$ at the resonance frequency. A feedback controller regulates the oscillation amplitude, $a$ of the Z-actuator by adjusting the amplitude, $I_d$ of the current driving the piezoelectric transducers. Specifically, the amplitude, $I_d$ of the drive current, generated by the high-voltage amplifier at the drive frequency $f_d = 39.59 \, \text{kHz}$, is measured using a lock-in amplifier (UHFLI, Zurich Instruments) and further regulated using a Proportional-Integral (PI) controller by varying the drive voltage, $V_d$ applied to the piezoelectric transducers. The regulation of the drive current eliminates non-linear Z-motion due to hysteresis of the piezoelectric transducers.  

The planar coil, which consists of four zones, was designed on a 3-layer Printed Circuit Board (PCB). Zones I and III are located on the first layer, while Zones II and IV are designed in the second layer, which is placed $140\, \mu\text{m}$ below the first layer. To prevent short-circuiting between the tracks, the long parallel segments of the second track in each zone are connected using vias passing through the third layer. Each track consists of 22 parallel segments, each with a length of $74\, \text{mm}$, a width of $127\, \mu\text{m}$, and a thickness of $35\, \mu\text{m}$. Although the parallel segments forming the planar coil tracks have finite length, and each track comprises only 22 such segments, unlike the idealized assumptions of infinite length and count in the analytical model, the predicted electromagnetic field, force, and stiffness closely agree with those obtained from finite element analysis (FEA) (see Supplementary Note~2).

The current through the tracks is controlled using voltage-controlled current sources based on high-current operational amplifiers (OPA549, TI). The drive voltages to these current sources are regulated using a prototype controller (MicroLabbox, DSpace), programmed in Simulink and MATLAB. Finally, a copper plate with a thickness of $600\, \mu\text{m}$ was integrated above the PCB to damp the in-plane motion of the levitating magnetic plate through eddy current damping. The PCB was mounted on an aluminum platform with a through hole at its center. The minimum vertical separation between the top surfaces of the PCB and the Z-actuator is maintained at $q = 800\, \mu\text{m}$ to ensure that the levitating magnetic plate does not touch the copper sheet.

Two different position measurement systems were employed to monitor the 3D motion of the levitating magnetic plate. The Z-motion was measured using an optical interferometer (PICOSCALE, SmarAct), which offers a measurement bandwidth of up to \( 2.5 \, \text{MHz} \) and an RMS positioning resolution better than \( 1 \, \text{nm} \) (see Supplementary Note 3). Accurate Z-displacement measurements were possible due to the plate’s high vertical stability and large surface area, which precisely reflected the laser beam to the millimeter-scale optical sensor of the interferometer.

In contrast, measuring the XY-motion was more challenging because of the plate’s small thickness and significant lateral oscillations, particularly under low electromagnetic trapping stiffness and in the absence of eddy current damping. Therefore, in-plane motion was measured using an in-house developed system based on sub-pixel digital image cross-correlation technique (see Supplementary Note 3). This setup includes a digital camera (MC031MG-SY-UB, Ximea) mounted on a modular microscope (Olympus, BXFM) for image acquisition. Two objective lenses with optical magnifications of 5 (LMPLFLN5X, Olympus) and 20 (LMPLFLN20X, Olympus) were used for large-range and high-resolution measurements, respectively. The system supports an in-plane motion range of \( 1.42 \, \text{mm} \), an RMS positioning resolution of \( 2 \, \text{nm} \), and a maximum measurement bandwidth of \( 1 \, \text{kHz} \).

Sections 4.1 and 4.2 discuss the characterization of the out-of-plane and in-plane motion of the developed positioner, respectively.

\subsection{Characterization of the Z-Motion}

In this section, we experimentally characterize the motion range, positioning resolution, and bandwidth along the Z-axis. The vertical motion range, \( \Delta h \), is defined as the difference between the maximum and minimum levitation heights of the magnetic plate, and it depends on the oscillation amplitude \( a \) of the Z-actuator and the total current \( I_x + I_y \) through the coil tracks. To estimate the levitation height \( h \), we solve the force balance equation along the Z-axis (as described in Section 3.2) under the simplifying assumptions \( h/p = 0 \) and \( \xi = 0.8 \). This provides the approximate expression \( h \approx 9.57a / \sqrt{1 + 0.28(I_x + I_y)} \), which indicates that the levitation height \( h \) can be increased either by increasing the oscillation amplitude \( a \) or by reducing the total current \( I_x + I_y \) through the coil tracks.

\begin{figure}[hbt!]
    \begin{centering}
  \includegraphics[width=0.42\paperwidth]{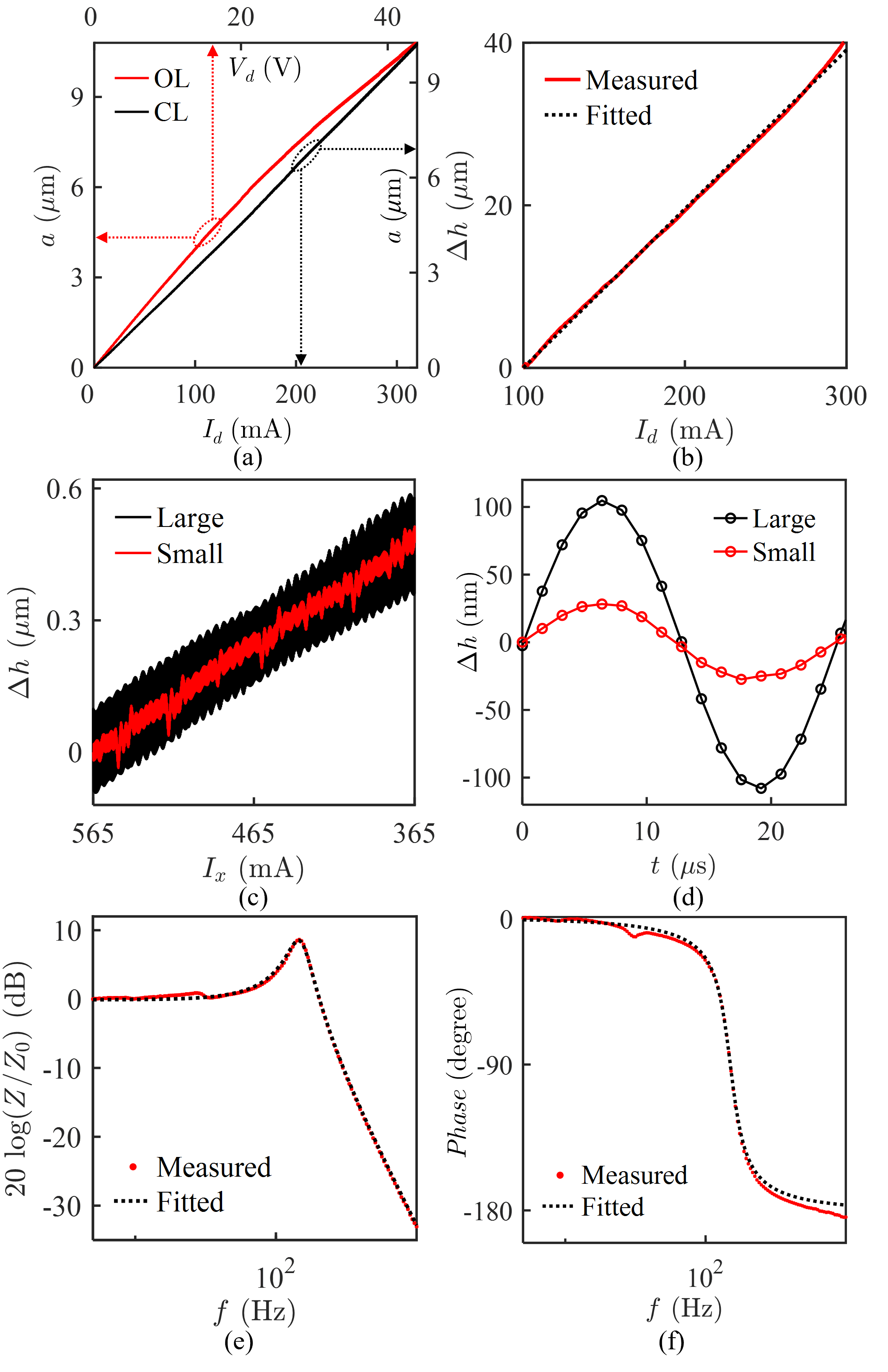}
    \par
    \end{centering}

\caption{\small 
Plots showing: (a) Variation in the oscillation amplitude \( a \) of the Z-actuator as a function of the piezo-drive current \( I_d \), where the red and black curves represent open-loop and closed-loop operation, respectively; (b) Large-range Z-displacement of the levitating magnetic plate achieved using acoustic Z-forces; (c) Small-range, high-resolution Z-displacement enabled by electromagnetic Z-forces, where the red and black plots correspond to the small and large magnetic plates, respectively; (d) Small-amplitude, high-frequency Z-oscillations of the levitating plates, with red and black curves representing the small and large plates, respectively; (e–f) Frequency response of the levitating magnetic plate along the Z-axis. The fitted curves for large-range Z-motion and frequency response are shown as dotted lines, while the fitted curves for high-frequency Z-oscillations are shown as solid lines.
}

    \label{Z-motion}
    \vspace{-4pt} 
\end{figure}

We first demonstrate levitation height control by varying the oscillation amplitude \( a \). The amplitude is adjusted by changing the piezo-drive voltage \( V_d \) from 0 to \( 42\, \text{V} \), while maintaining a constant drive frequency of \( f_d = 39.59\, \text{kHz} \). The resulting variation in \( a \) is measured and shown as the red plot in Fig.~\ref{Z-motion}(a). The plot reveals that the amplitude increases nonlinearly with \( V_d \), primarily due to hysteresis in the piezoelectric transducers driving the mechanical horn. To eliminate this nonlinearity, a proportional-integral (PI) controller \( C(s) = 1 + 100/s \) is implemented to regulate the piezo drive current \( I_d \) by controlling the drive voltage \( V_d \). The black plot in Fig.~\ref{Z-motion}(a) shows that, with feedback control, the amplitude \( a \) increases linearly from 0 to \( 10.25\, \mu\text{m} \) as \( I_d \) increases from 0 to \( 320\, \text{mA} \). The maximum oscillation amplitude is limited to \( 10.25\, \mu\text{m} \) due to the \( 40\, \text{V} \) output voltage range of the electronics used to drive the piezo transducers.
Next, the magnetic plate is positioned above the stationary Z-actuator. A fixed current of \( 400\, \text{mA} \) is applied to all PCB tracks to stabilize the plate horizontally at \( x_o = y_o = 0.3175\, \text{mm} \). Subsequently, the piezo-drive frequency \( f_d \) is reduced to \( 39.56\, \text{kHz} \), corresponding to the resonance frequency of the coupled Z-actuator and levitating plate. The minimum piezo-drive current \( I_d \) required to achieve levitation is experimentally determined to be approximately \( 100\, \text{mA} \), corresponding to a minimum levitation height of about \( 26\, \mu\text{m} \).

Finally, to determine the vertical motion range, the piezo-drive current \( I_d \) was varied linearly from \( 100\, \text{mA} \) to \( 300\, \text{mA} \). The resulting variation in mean levitation height, measured using the interferometer, \( \Delta h \), is shown as the solid red line in Fig.~\ref{Z-motion}(b), while a linear model obtained via least-squares fitting is represented by the dotted black line. The plots confirm that \( h \) increases linearly with \( a \), consistent with the simplified analytical expression for the levitation height. However, the experimentally measured displacement, \( \Delta h = 40\, \mu\text{m} \), is approximately \( 25\% \) lower than the analytically predicted value of \( 53.3\, \mu\text{m} \). This discrepancy is primarily attributed to deviations between the assumptions in the analytical model and the actual physical implementation. In particular, the model presented in \cite{wang2021stiffness} assumes a lightweight levitating plate such that the squeeze ratio \( a/h \leq 0.05 \). In contrast, the developed positioner uses a heavier magnetic plate, resulting in a squeeze ratio \( a/h \) that is at least twice that used in the model. This increased squeeze ratio leads to a reduced levitation height, explaining the observed deviation. It is worth noting that the vertical motion range can be further enhanced by increasing the drive current beyond \( 300\, \text{mA} \), provided the drive electronics support higher actuation voltages and currents.

Instead of varying the oscillation amplitude, the simplified levitation height expression discussed earlier also suggests that the levitation height can be adjusted by modifying the currents through the electromagnetic coil. The black plot in Fig.~\ref{Z-motion}(c) shows that the mean levitation height of the plate increases linearly by approximately \( 0.5\, \mu\text{m} \) as \( I_x \) is reduced from \( 565\, \text{mA} \) to \( 365\, \text{mA} \), while both the oscillation amplitude and the currents through the other tracks are held constant at \( a = 3.2\, \mu\text{m} \) and \( I_y = 565\, \text{mA} \), respectively. Using the simplified analytical expression, the expected displacement was calculated to be \( 0.59\, \mu\text{m} \), which closely matches the experimental result. It is important to note that the magnitude of the electromagnetic Z-forces is significantly smaller than that of the acoustic Z-forces, thereby limiting the achievable vertical displacement through current modulation alone to only a few microns.

In addition to the variation in mean levitation height, the black plot in Fig.~\ref{Z-motion}(c-d) shows small-amplitude oscillations of the levitating plate at the drive frequency, \( f_d = 39.56 \, \text{kHz} \). These high-frequency oscillations are induced by dynamic acoustic forces acting on the magnetic plate \cite{wang2021stiffness}. For the larger magnetic plate shown in Fig.~\ref{Positioner}, these oscillations limit the root-mean-square (RMS) positioning resolution along the Z-axis to \( 74 \, \text{nm} \). To reduce the amplitude of these high-frequency oscillations, we designed a smaller square plate with an edge length of \( 25.2 \, \text{mm} \) and the same thickness of \( 1.5 \, \text{mm} \), embedded with a checkerboard magnetic array consisting of 132 magnets. Compared to the larger plate, the smaller plate exhibits mechanical Z-resonance modes that are located farther from the drive frequency \( f_d \), resulting in more than a threefold improvement in the RMS positioning resolution, as shown by the red plot in Fig.~\ref{Z-motion}(c-d). Fig.~\ref{Z-motion}(d) also provides a zoomed-in view of the high-frequency oscillations over one time period for both plates. The oscillation amplitudes for the larger and smaller plates were measured to be \( 104.7 \, \text{nm} \) and \( 28.2 \, \text{nm} \), respectively. It should be noted that the smaller plate was used solely to demonstrate the potential improvement in Z-axis positioning resolution achieved by reducing high-frequency oscillations through mechanical design optimization. To ensure consistency, all other experiments were conducted using a single plate, namely, the larger magnetic plate shown in Fig.~\ref{Positioner}.

Finally, to characterize the dynamic response along the Z-axis, we measured the frequency response of the levitating magnetic plate. A sinusoidal electromagnetic force \( F_{zy} \) with small amplitude and linearly varying frequency from \( 5\, \text{Hz} \) to \( 1\, \text{kHz} \) was applied to the plate, while the acoustic levitation force was held constant. The resulting small-amplitude Z-motion was measured using the interferometer and fed into the lock-in amplifier, which extracted the oscillation amplitude and phase at each excitation frequency \( f \). Specifically, the excitation currents were applied as \( I_{1x} = I_{2x} = 0.4\, \text{A} + 0.1 \sin(2\pi f t)\, \text{A} \), \( I_{1y} = I_{2y} = 0.4\, \text{A} \), with the piezo-drive current maintained at \( I_d = 0.1\, \text{A} \). The normalized magnitude and phase responses of the Z-motion are shown in Fig.~\ref{Z-motion}(e-f). A second-order model was fitted to the measured data, from which the natural frequency, damping ratio, and 3-dB bandwidth were estimated to be \( 150.11\, \text{Hz} \), \( 0.19 \), and \( 171\, \text{Hz} \), respectively. The analytically predicted natural frequency, calculated using \( \frac{1}{2\pi} \sqrt{k_z / m_p} = 152\, \text{Hz} \), where \( k_z = k_a - (k_x+k_y) \), closely matches the experimentally obtained value. The measured and fitted responses are shown in red and black, respectively, in Fig.~\ref{Z-motion}(e-f).  

\vspace{-6pt} 

\subsection{Characterization of XY-Motion}

In this section, we experimentally demonstrate three key enhancements in horizontal positioning performance resulting from eddy current damping and electromagnetic trapping in the XY-plane. First, we quantify the increase in XY-plane damping due to eddy currents induced in the copper sheet. Second, we demonstrate the enhancement in XY-stiffness resulting from electromagnetic trapping. Third, we showcase the ability to apply controlled horizontal electromagnetic forces to the levitating magnetic plate. These three improvements collectively enable precision XY-positioning with a millimeter-scale motion range, nanometer-scale resolution, and a positioning bandwidth exceeding tens of hertz. Throughout the characterization experiments, the levitation height of the magnetic plate was maintained at \( h \approx 26\, \mu\text{m} \) by regulating the piezo-drive current to \( I_d = 0.1\, \text{A} \).

Since the levitating magnetic plate at its stable equilibrium position \( (x_o, y_o, z_o) \) can be modeled as a mass–spring–damper system (as discussed in Section~3.2), we use experimentally determined damping ratios and spring constants to demonstrate the improvements achieved through eddy current damping and electromagnetic trapping, respectively. To stabilize the magnetic platform at \( x_o = y_o = 0 \), equal currents were applied to the first track of all four zones, while zero current was applied to the second track of each zone. Subsequently, small step and sinusoidal currents were applied to the second tracks of all zones to generate horizontal actuation forces in the form of step and sinusoidal inputs. The resulting dynamic motion of the levitating platform was measured using the in-plane position measurement system described earlier. The damping ratio and spring constant were then extracted from the measured motion profiles to quantify the improvements. Owing to symmetry in electromagnetic trapping stiffness and actuation forces, the performance along the X- and Y-axes is expected to be identical. Therefore, in the following analysis, we focus on characterizing the positioning performance along a single axis, namely the X-axis.

\begin{figure}[hbt!]
    \begin{centering}
  \includegraphics[width=0.42\paperwidth]{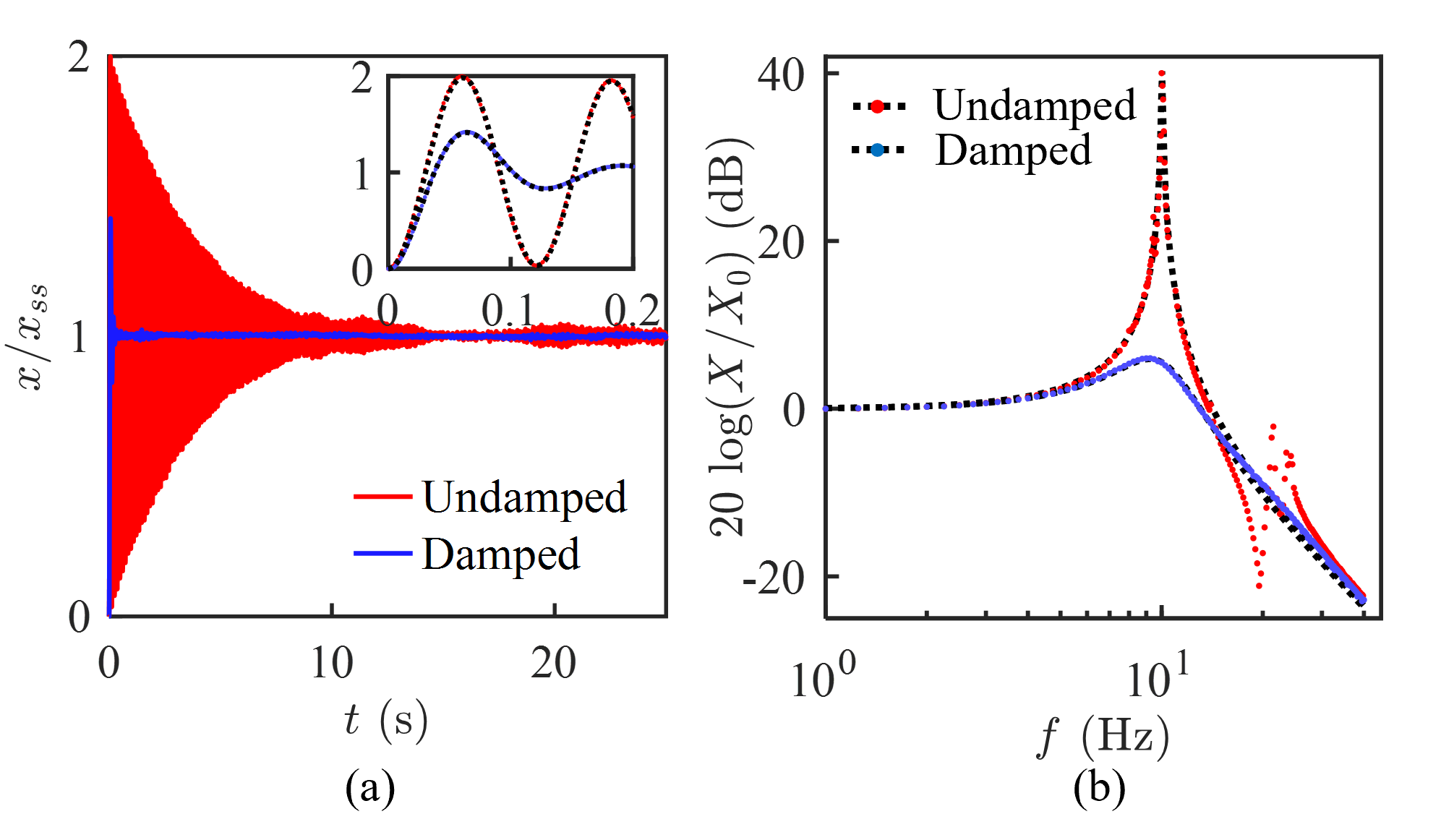}
    \par
    \end{centering}

\caption{\small 
Plots showing: (a) Step responses and (b) Frequency responses of the system with and without eddy current damping. In both plots, the red curves represent the undamped response, while the blue curves represent the response with eddy current damping. The fitted data are shown as dotted black lines. In the step response, the settling time is reduced from 14~s to 0.3~s and the peak overshoot is reduced from 100\% to 40\% due to eddy current damping. In the frequency response, the resonant peak magnitude is attenuated by a factor of 52 with damping.
}
\label{damping}
\vspace{-8pt} 
\end{figure}

First, we report the improvement in the damping ratio resulting from eddy current induction. A constant current of \( I_{1x} = I_{1y} = 0.4\, \text{A} \) was applied to electromagnetically stabilize the levitating magnetic plate. Then a step actuation force of \( F_{x\text{act}} = 0.14\, \text{mN} \) is applied along the X-axis by varying the current \( I_{2x} \) from 0 to \( 0.01\, \text{A} \). The normalized step responses for the undamped and damped systems are plotted in Fig.~\ref{damping}(a). A nonlinear model, fitted to the measured data using the least-squares method, was used to estimate the natural frequency and damping ratio. The fitted models are shown as dotted lines in the inset of Fig.~\ref{damping}(a), where the close agreement with the measured response shows the accuracy of the dynamic model derived in Section 3.2. The estimated damping ratios for the undamped and damped systems are 0.005 and 0.27, respectively, corresponding to a 54-fold increase in damping due to eddy currents. In addition, the increased damping improves the RMS positioning resolution by approximately a factor of four, as indicated by the steady-state responses. The natural frequencies of the damped and undamped systems were estimated to be \( 8.1\, \text{Hz} \) and \( 8.2\, \text{Hz} \), respectively, indicating that the eddy currents do not affect the trapping stiffness.

Additionally, the X-frequency response of the positioner, with and without eddy current damping, was measured for comparison. A constant current of \( I_{1x} = 0.6\, \text{A} \) was used for stabilization, while the second track current \( I_{2x} \) was used to apply a sinusoidal actuation force to the levitating plate. Specifically, a current of the form \( I_{2x} = 0.01 \sin(2\pi f t)\, \text{A} \) was applied, where the excitation frequency \( f \) was varied from \( 1\, \text{Hz} \) to \( 40\, \text{Hz} \) to generate the actuation force \( F_{x\text{act}} \) required for frequency response analysis. The normalized magnitude responses of the damped and undamped systems are shown in Fig.~\ref{damping}(b). The oscillation amplitude at each excitation frequency was estimated offline using a custom lock-in detection algorithm. A nonlinear model was then fitted to the experimental data using the least-squares method to estimate the natural frequency and damping ratio. The fitted models are represented as dotted lines in Fig.~\ref{damping}(b). The estimated damping ratios for the undamped and damped systems were 0.005 and 0.261, respectively, indicating a 52.2-fold increase in damping due to eddy currents. The estimated natural frequency for both cases was approximately \( 10\, \text{Hz} \), confirming that eddy current damping does not affect the trapping stiffness. These results, consistent with the step-response analysis, confirm that the induced eddy currents significantly improve damping in the XY-plane. Additionally, the frequency response plots show that the increased damping effectively suppresses unwanted oscillation modes around \( 20\, \text{Hz} \), which are likely caused by cross-coupling between in-plane and out-of-plane motion in the undamped system.

\begin{figure}[hbt!]
    \begin{centering}
   \includegraphics[width=0.42\paperwidth]{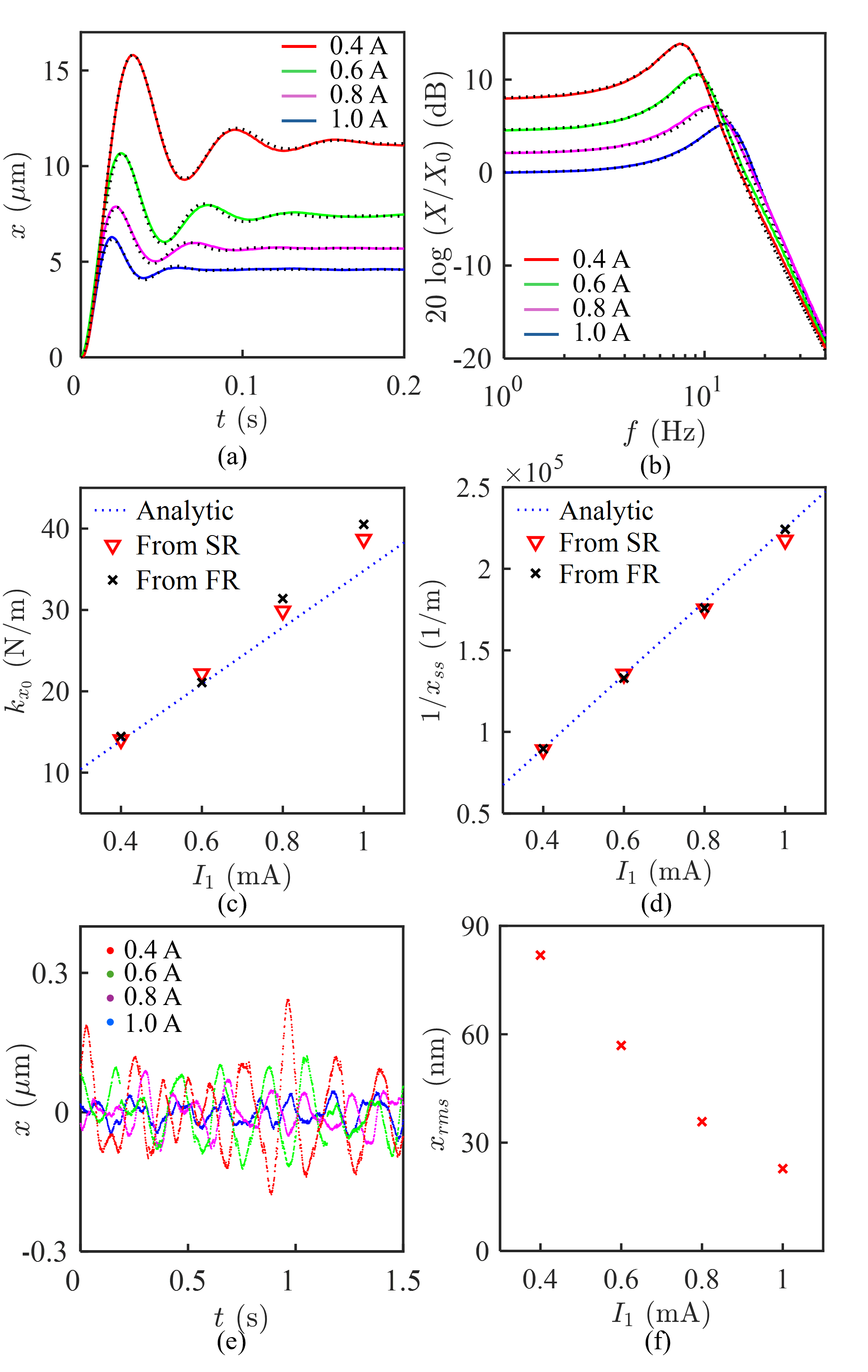}
    \par
    \end{centering}

\caption{\small 
Plots showing: (a) Step responses for different track currents \( I_{1x} \); (b) Frequency responses for different track currents \( I_{1x} \); (c) Trapping stiffness \( k_{xo} \) as a function of track current \( I_{1x} \); (d) Reciprocal of the steady-state displacement \( 1/x_{ss} \) and the low-frequency oscillation amplitude \( 1/X_{0} \) as functions of \( I_{1x} \); (e) Oscillation of the positioner from its steady-state position without actuation; and (f) RMS positioning resolution, each evaluated for varying track current \( I_{1x} \). The trapping stiffness and steady-state displacement were derived from both step and frequency responses. The symbols \( \nabla \) and \( \times \) represent values estimated from step and frequency responses, respectively. Fitted and analytical results are shown as dotted lines. The plots demonstrate that the trapping stiffness increases linearly with the current, thereby reducing the amplitude of undesired horizontal oscillations of the levitating magnetic platform.
}

    \label{StiffnessvsI}
    \vspace{-8pt} 
 \end{figure}

Next, experiments were conducted to evaluate the enhancement in X-axis trapping stiffness, \( k_{xo} \), resulting from the applied track current \( I_{1x} \). According to Equation~\ref{Fxkx}, the X-trapping stiffness for the hybrid positioner is given by \( k_{xo} = 190.7\, \phi\, I_{1x}\, \text{N/(A.m)} \), where \( \phi \) is a parameter that depends on the Z-position of the magnetic plate. The X-trapping stiffness was experimentally determined using both step and frequency response analyses. Four values of the trapping current: \( 0.4\, \text{A} \), \( 0.6\, \text{A} \), \( 0.8\, \text{A} \), and \( 1.0\, \text{A} \) were selected to investigate the effect of current magnitude on trapping stiffness while keeping Z-position constant at its minimum value of \( z = 1076\, \mu\text{m} \). For the specified Z-position, the parameter \( \phi = 0.18\) and the electromagnetic stiffness is given by \( k_{xo} = 34.81\, I_{1x}\, \text{N/(A.m)} \).  For each value of \( I_{1x} \), step and frequency responses were obtained by varying \( I_{2x} \) as described previously. The corresponding experimental responses are shown as solid lines in Fig.~\ref{StiffnessvsI}(a) and (b), while the least-squares fitted models are shown as dotted lines. The stiffness values were estimated by multiplying the square of the experimentally determined natural frequency with the mass of the levitating plate, \( m_p \). These experimentally obtained stiffness values closely match the analytical values of \( k_{xo} \), as shown in Fig.~\ref{StiffnessvsI}(c). The close agreement validates the electromagnetic model described in Section~3. The plot in Fig.~\ref{StiffnessvsI}(c) also confirms that the trapping stiffness \( k_{xo} \) increases linearly with the applied current \( I_{1x} \). Notably, the use of in-plane electromagnetic forces allows the horizontal trapping stiffness to reach values as high as \( 40\, \text{N/m} \), which is three orders of magnitude greater than the previously reported maximum of \( 0.04\, \text{N/m} \) achievable with acoustic in-plane restoring forces \cite{aono2019increase,li2024contact}.

Although the trapping stiffness was primarily estimated using the natural frequency, it can also be determined by dividing the actuation force \( F_{x\text{act}} \) by either the steady-state displacement \( x_{\text{ss}} \) obtained from the step response or the low-frequency sinusoidal oscillation amplitude \( X_0 \) obtained from the frequency response. Since the same actuation current \( I_{2x} = 0.01 \, \text{A} \ \) was used throughout, the actuation force remained constant at \( F_{x\text{act}} = 0.14\, \text{mN} \). Therefore, both the reciprocal of the steady-state displacement \( 1/x_{\text{ss}} = k_{xo}/F_{x\text{act}} \) and the reciprocal of the low-frequency oscillation amplitude \( 1/X_0 = k_{xo}/F_{x\text{act}} \) are also expected to vary linearly with \( I_{1x} \) as the trapping stiffness \( k_{xo} \) increases linearly with the trapping current \( I_{1x} \). This linear trend is confirmed experimentally, as shown in Fig.~\ref{StiffnessvsI}(d). The analytical values of \( k_{xo}/F_{x\text{act}} \), represented by the dotted blue line, closely match the experimental data points.

Finally, we measured the X-position of the levitating plate for different trapping stiffness values, without applying any external actuation force. The plot in Fig.~\ref{StiffnessvsI}(e) shows that the amplitude of undesired X-motion decreases as the trapping stiffness increases. These residual vibrations are primarily caused by fluctuations in the in-plane acoustic forces generated during near-field acoustic levitation. The RMS positioning resolution was calculated for each trapping current and is plotted in Fig.~\ref{StiffnessvsI}(f). The results show nearly a fourfold improvement in positioning resolution from \( 82\, \text{nm} \) to \( 23\, \text{nm} \), corresponding to a 2.5-fold increase in trapping stiffness. The damping ratio remained nearly constant at 0.27, consistent with the fact that it is independent of the trapping current. It is noteworthy that the combination of electromagnetic trapping and eddy current damping improved the positioning resolution from the millimeter scale to the nanometer scale, representing a three to four orders of magnitude enhancement \cite{gabai2019contactless,kikuchi2021development}.

After demonstrating the linear dependence of trapping stiffness \( k_{xo} \) on the trapping current \( I_{1x} \), experiments were conducted to investigate its variation with the parameter 
$
\phi = \text{sech}(2\pi(\tilde{z} - \tilde{t}/2)) - \text{sech}(2\pi(\tilde{z} + \tilde{t}/2)),
$ where \( \tilde{z} = z/p \) and \( \tilde{t} = t/p \) represent the normalized Z-position and magnet thickness, respectively. While \( \phi \) is a design parameter that typically remains fixed during operation, this experiment explores how changes in \( \phi \) affect positioning performance and compares the results with the derived analytical model. The parameter \( \phi \) was varied by adjusting the distance \( z = h + q + t_s/2 \) between the levitating magnetic plate and the PCB (see Fig.~\ref{modeling}(a)). This was accomplished by changing \( q \), the vertical distance between the top surface of the stationary Z-actuator and the PCB, using a precision micrometer stage. The piezo-drive current was held constant at \( I_d = 0.1\, \text{A} \) to maintain a constant mean levitation height of \( h \approx 26\, \mu\text{m} \), ensuring that acoustic dynamics remained unchanged throughout the experiment. Five values of \( q \) were selected, ranging from \( 800\, \mu\text{m} \) to \( 1200\, \mu\text{m} \), in increments of \( 100\, \mu\text{m} \), resulting in corresponding values of \( z \) from \( 1076\, \mu\text{m} \) to \( 1476\, \mu\text{m} \). The minimum value of \( q \) was selected to prevent contact between the levitating plate and the copper sheet. It is important to note that changes in \( q \) also alter the distance between the magnetic plate and the copper sheet, given by \( h + q-t_s \), thus affecting the eddy current damping.

\begin{figure}[hbt!]
    \begin{centering}
   \includegraphics[width=0.42\paperwidth]{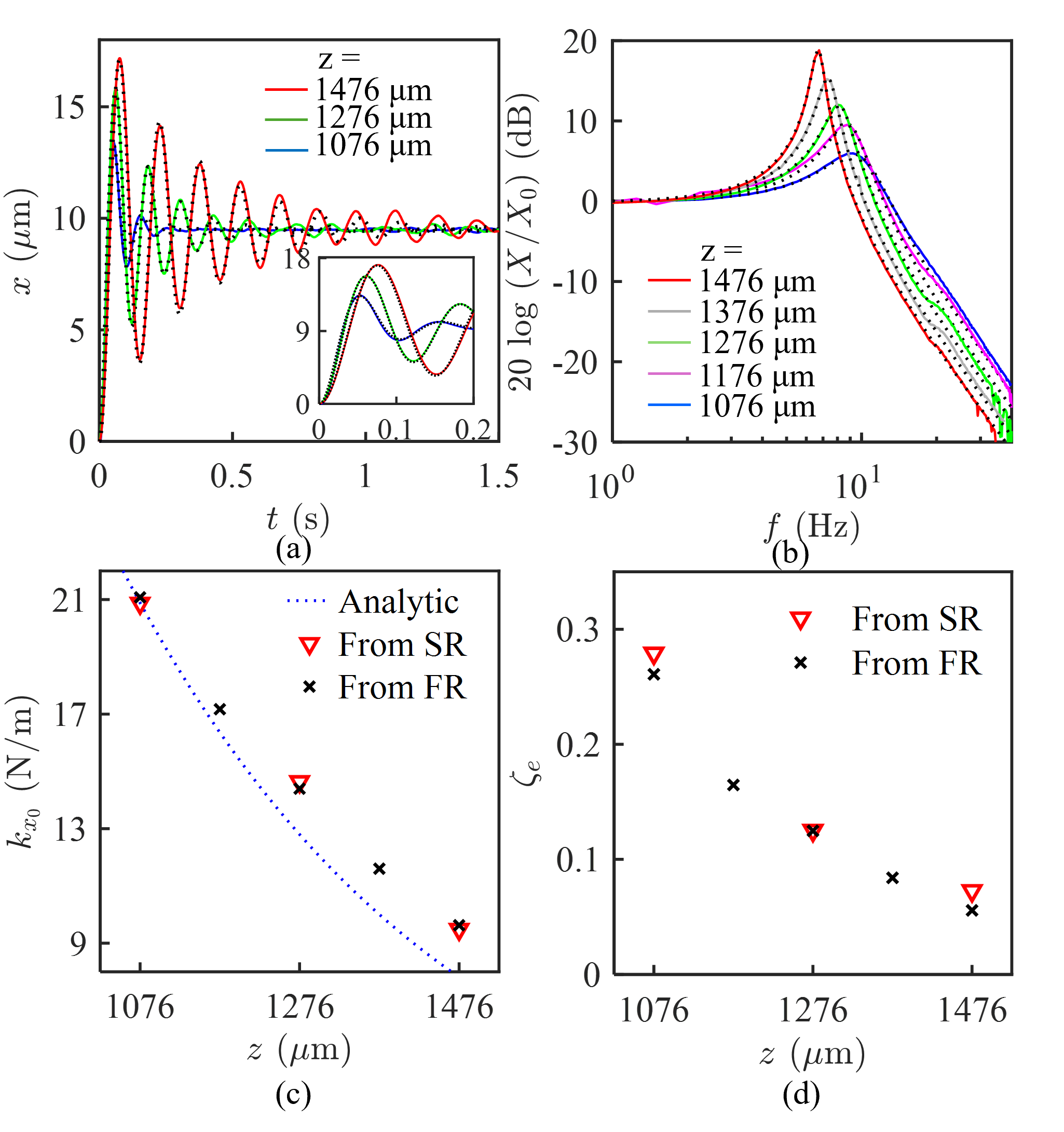}
    \par
    \end{centering}
\caption{\small
Plots showing: (a) Step responses; (b) Frequency responses; (c) Trapping stiffness \( k_{xo} \); and (d) Eddy current damping factor, each as a function of the separation distance \( z \) between the magnetic plate and the PCB. The stiffness and damping factor were extracted from both step and frequency response data. The symbols \( \nabla \) and \( \times \) represent values estimated from step and frequency responses, respectively. Fitted results are shown as dotted black lines, while the analytical values of trapping stiffness from the derived model are shown as a dotted blue line. The plots illustrate that increasing the separation \( z \) reduces both the electromagnetic trapping stiffness and the eddy current damping.
}
    \label{Stiffnessvsphi}
    \vspace{-8 pt} 
 \end{figure}

Both step and frequency responses were obtained to evaluate the variation in trapping stiffness \( k_{xo} \) and damping ratio \( \zeta_e = b_e / (2\sqrt{m_p k_{xo}}) \) as a function of the vertical distance \( z \). During these experiments, the levitating magnetic plate was stabilized using a fixed trapping current of \( I_{1x} = 0.6\, \text{A} \). To generate the step response, the actuation current \( I_{2x} \) was varied from \( 0 \) to \( 0.01\, \text{A} \), resulting in an actuation force \( F_{x\text{act}} = 0.77 \, \phi \, \text{mN} \). The corresponding step responses are shown in Fig.~\ref{Stiffnessvsphi}(a). For the frequency response, a sinusoidal current \( I_{2x} = 0.01 \sin(2\pi f t)\, \text{A} \) was applied, where \( f \) was swept from \( 1\, \text{Hz} \) to \( 40\, \text{Hz} \). The resulting frequency responses are shown in Fig.~\ref{Stiffnessvsphi}(b). The stiffness values estimated from both the step and frequency responses are plotted in Fig.~\ref{Stiffnessvsphi}(c), alongside analytical predictions using \( k_{xo} = 114.4 \phi\, \text{N/m} \). The close agreement between experimental and analytical values confirms that the model described in Section 3 accurately captures the variation in trapping stiffness with the Z-position of the magnetic plate. As expected, the trapping stiffness decreases with increasing separation distance \( z \) between the current-carrying tracks and the levitating magnetic platform. Additionally, Fig.~\ref{Stiffnessvsphi}(d) shows the corresponding reduction in eddy current damping as the distance between the magnets and the copper sheet increases. Notably, the steady-state displacement \( x_{\text{ss}} = F_{x\text{act}} / k_{xo} \) obtained from the step response and the low-frequency oscillation amplitude \( X_0 = F_{x\text{act}} / k_{xo} \) obtained from the frequency response remain nearly constant with varying \( z \). This invariance occurs because both the actuation force and the trapping stiffness scale proportionally with the factor \( \phi \) as \( z \) increases.

\begin{figure}[hbt!]
    \begin{centering}
   \includegraphics[width=0.43\paperwidth]{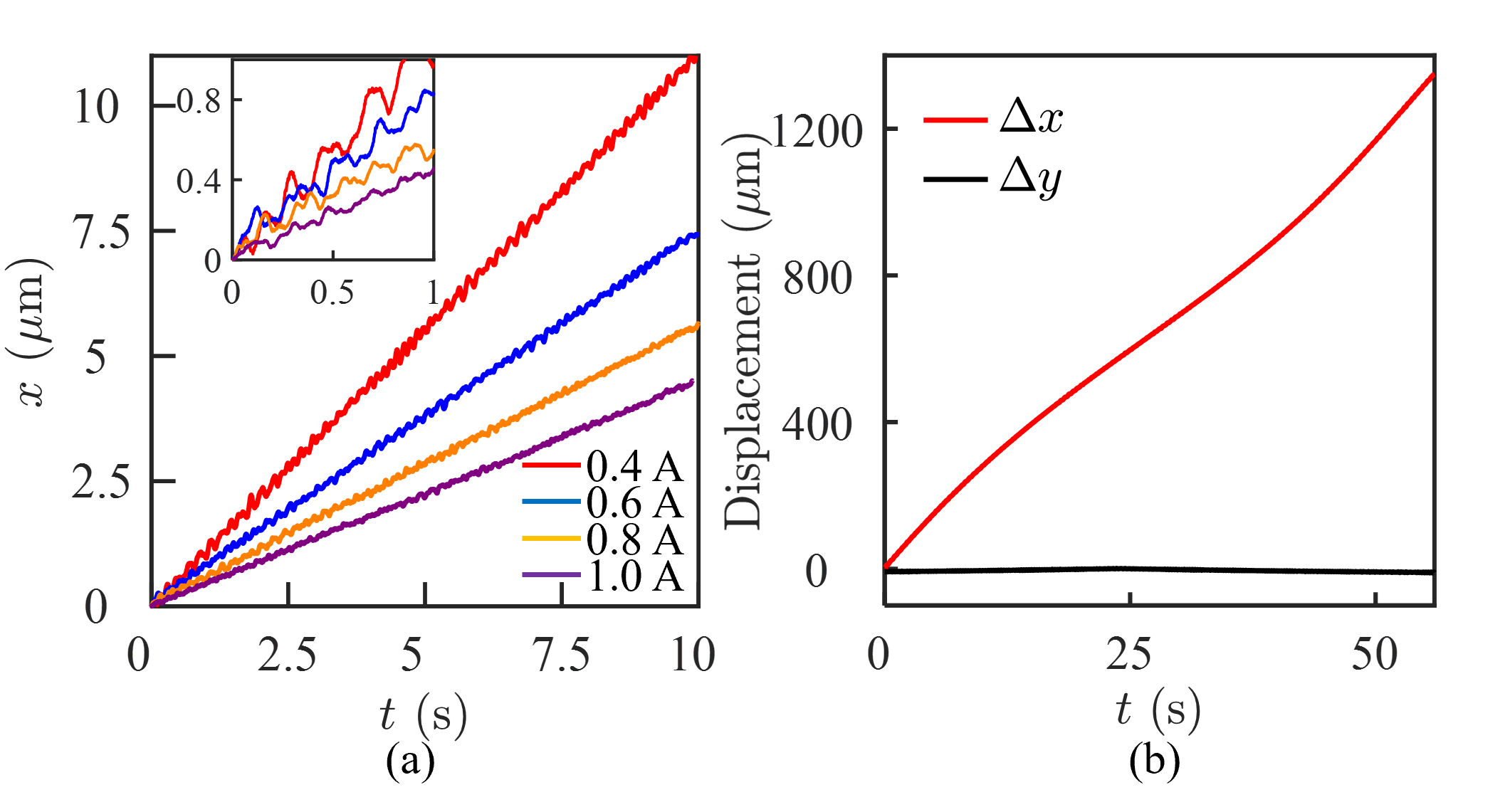}
    \par
    \end{centering}
\caption{\small
Plots showing: (a) Small-range, high-resolution linear X-motion obtained by keeping \( I_{1x} \) constant and linearly varying \( I_{2x} \); the plots also illustrate the improvement in positioning resolution with increasing trapping current \( I_{1x} \); (b) Large-range linear X-motion with negligible cross-axis Y-motion, achieved using sinusoidal actuation currents \( I_{1x} \) and \( I_{2x} \) that are \( 90^\degree \) out of phase, resulting in uniform trapping stiffness throughout the motion range.
}
    \label{X-motion}
 \end{figure}

Finally, we demonstrate precision positioning along the X-axis by increasing damping and trapping stiffness. The separation between the magnets and the PCB was reduced to its minimum value, \( z = 1076\, \mu\text{m} \), thus enhancing both eddy current damping and electromagnetic trapping stiffness as discussed in the previous paragraph. To investigate the effect of trapping stiffness on positioning resolution, four values of the trapping current \( I_{1x} \) were tested: \( 0.4\, \text{A} \), \( 0.6\, \text{A} \), \( 0.8\, \text{A} \), and \( 1\, \text{A} \). Subsequently, the current \( I_{2x} \) was varied to apply the actuation force \( F_{x\text{act}} = 0.014 I_{2x} \, \text{N} \), enabling precise positioning of the levitating platform along the X-axis. As shown in Fig.~\ref{X-motion}(a), the displacement of the levitating plate varied linearly with the actuation current \( I_{2x} \). The experimentally measured displacements closely match the analytical predictions obtained using \( \Delta x = F_{x\text{act}} / k_{xo} \). Furthermore, the results show that the RMS positioning resolution improved from \( 80\, \text{nm} \) to \( 20\, \text{nm} \) as the trapping current \( I_{1x} \) increased from \( 0.4\, \text{A} \) to \( 1\, \text{A} \). These results highlight the capability of the positioner to provide nanometer-scale positioning precision in the XY-plane, which was previously limited to sub-millimeter scale.  However, this increase in trapping stiffness also led to a decrease in the maximum motion range, which dropped from \( 11\, \mu\text{m} \) to \( 4.6\, \mu\text{m} \) across the tested current range.

Although the X-motion range can be enhanced by increasing the actuation current \( I_{2x} \), the maximum displacement is ultimately limited to \( \Delta x_o = \frac{p}{2\pi} \tan^{-1}\left( \frac{I_{2x}}{I_{1x}} \right) = 635\, \mu\text{m} \), assuming \( I_{2x} \gg I_{1x} \). Moreover, as the levitating magnetic plate moves away from the equilibrium position \( x_o = 0 \), the motion becomes increasingly nonlinear due to spatial variations in trapping stiffness. To achieve a larger X-motion range while maintaining a constant trapping stiffness \( k_{xo} \), sinusoidal actuation currents were applied: \( I_{1x} = 0.6 \cos(0.0628 t) \, \text{A} \) and \( I_{2x} = 0.6 \sin(0.0628 t) \, \text{A} \) as described in Section~3.2. This approach shifts the electromagnetic equilibrium position at a constant velocity of \( 25.4\, \mu\text{m/s} \), enabling the magnetic plate to follow the moving equilibrium point while maintaining a fixed trapping stiffness of \( k_{xo} = 21.4\, \text{N/m} \). The measured X- and Y-displacements of the levitating plate are shown in red and black, respectively, in Fig.~\ref{X-motion}(b). The results indicate a maximum X-motion range of \( 1.42\, \text{mm} \), limited by the range of the in-plane position measurement system, with negligible Y-displacement. The obtained results demonstrate the capability to provide millimeter scale motion range, nanometer scale positioning resolution with a positioning bandwidth of a few tens of hertz. The slight nonlinearity observed in the X-motion is attributed to deviations between the fabricated positioner and the idealized electromagnetic model discussed in Section~3.1. Specifically, the finite length and discrete number of conductors in the planar coil introduce higher-order harmonics into the magnetic field near the boundary of the planar coil, resulting in nonlinear motion effects that are absent in the infinite-coil model (see Supplementary Note 2). This nonlinearity could be mitigated by increasing the dimensions of the electromagnetic coil or by implementing active feedback control strategies. Additionally, the in-plane positioning bandwidth, which is currently limited to \( 16\, \text{Hz} \) for \( I_{1x} = 1\, \text{A} \) could potentially be improved by at least an order of magnitude through further optimization of the checkerboard magnetic array design \cite{10589479}. 

\section{Conclusion}

This paper presents a novel multi-axis positioner based on the hybrid combination of NFAL and steady current-based electromagnetic trap. Remarkably, this design enables 3D stabilization and precision positioning of the levitating platform without the need for feedback stabilization due to the complementary capabilities of the NFAL and electromagnetic trap. Here, we experimentally demonstrated in-plane linear motion, with a motion range of $1.42 \, \text{mm}$,  RMS positioning resolution of $20\, \text{nm}$, and a positioning bandwidth of $16\, \text{Hz}$. A travel range of $40\, \mu \text{m}$ was achieved along the Z-axis, with an RMS resolution of $20\, \text{nm}$ and a bandwidth of $171 \, \text{Hz}$. Compared to existing NFAL-based positioners, the proposed system provides a three orders of magnitude improvement in positioning resolution, at least a 10-fold increase in bandwidth in the XY-plane, and the added capability of multi-axis positioning. Future work will focus on enabling multi-axis rotational motion and further improvements in positioning resolution and bandwidth by enhancing the horizontal electromagnetic trapping stiffness. Additionally, the combination of NFAL with electromagnetic traps opens new avenues for research, such as NFAL within fluids and NFAL for micro-robotic applications.

\vspace{-2pt} 

\section*{Acknowledgment}
\begin{justify}
 The authors wish to thank Dr Hazhir Mahmoodi Nasrabadi, Apple Inc., for experimental assistance and Dr Pavan Kumar, Penn State, for discussion on NFAL. This research was supported by The University of Texas at Dallas through the James Von Ehr Distinguished Chair.
\end{justify}

\bibliographystyle{ieeetr}
\bibliography{citations}


\begin{IEEEbiography}[{\includegraphics[width=1in,height=1.25in,clip,keepaspectratio]{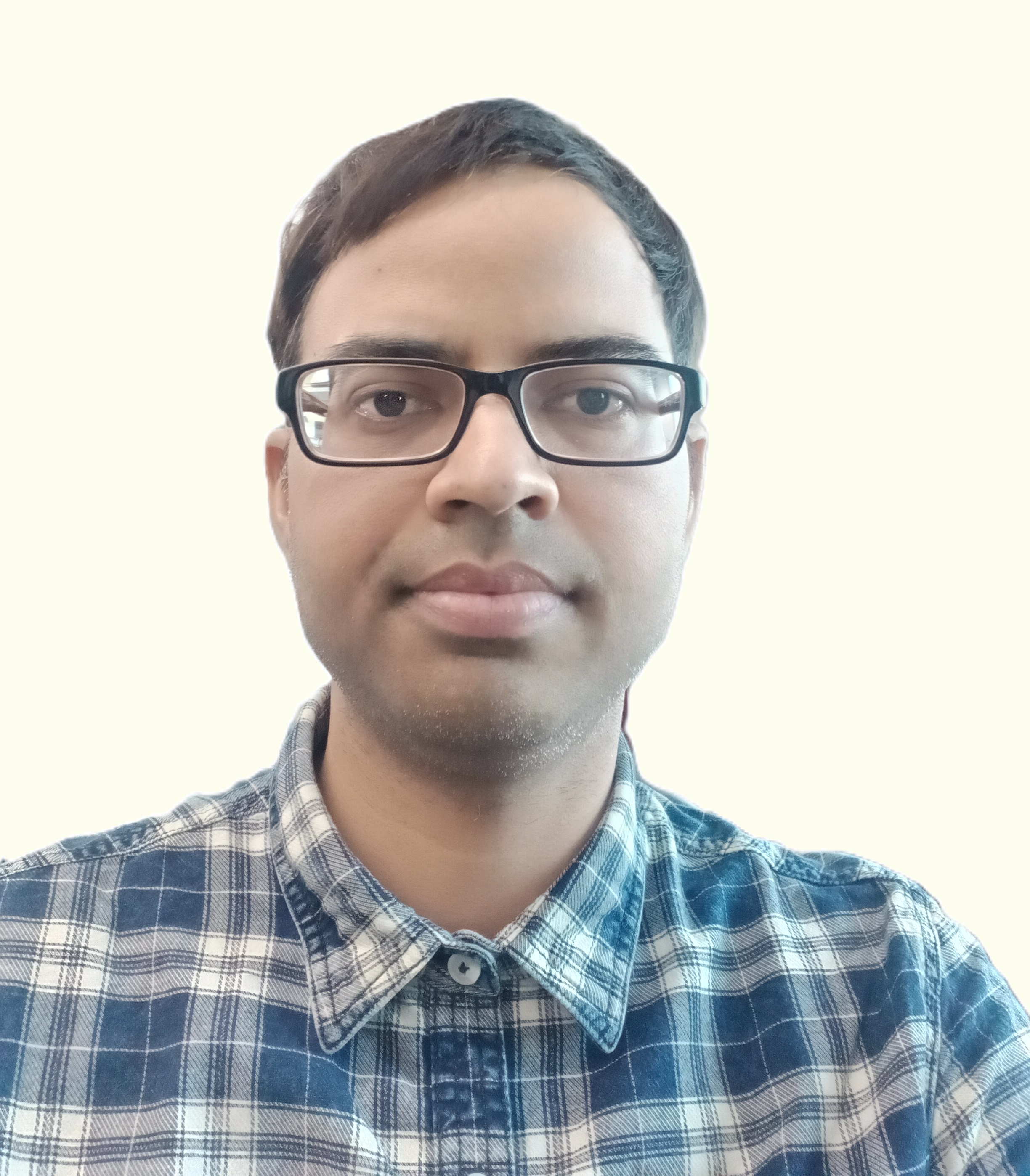}}]
{K.S. Vikrant} received his B.E. in Electronics and Instrumentation Engineering from the Institute of Technology and Management, Gwalior, in 2014 and both his MTech (Research) and PhD from the Indian Institute of Science, Bangalore in 2022. During his graduate research, he worked on precision actuation and measurement systems. He is currently a research scientist at the University of Texas at Dallas and is a member of the Laboratory for Dynamics and Control of Nanosystems. His current research interests include scanning probe microscopes, precision mechatronics, passive levitation-based micro and nanopositioners, and micro-robotics.
\end{IEEEbiography}
\begin{IEEEbiography}[{\includegraphics[width=1in,height=1.25in,clip,keepaspectratio]{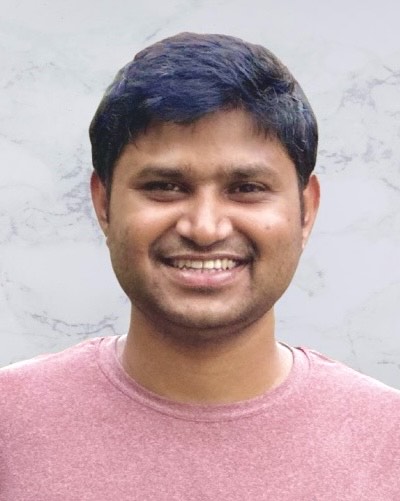}}]
{Prosanto Biswas} received his B.Sc. in Electrical and Electronic Engineering from the Bangladesh University of Engineering and Technology, Dhaka, in 2017. He completed his M.Sc. in Electrical Engineering at the University of Texas Rio Grande Valley in 2021. Currently, he is pursuing a Ph.D. at the University of Texas at Dallas and is a member of the Laboratory for Dynamics and Control of Nanosystems. His research focuses on precision mechatronics, nanofabrication, scanning probe microscopy and levitation-based nanopositioners.
\end{IEEEbiography}
\begin{IEEEbiography}[{\includegraphics[width=1in,height=1.25in,clip,keepaspectratio]{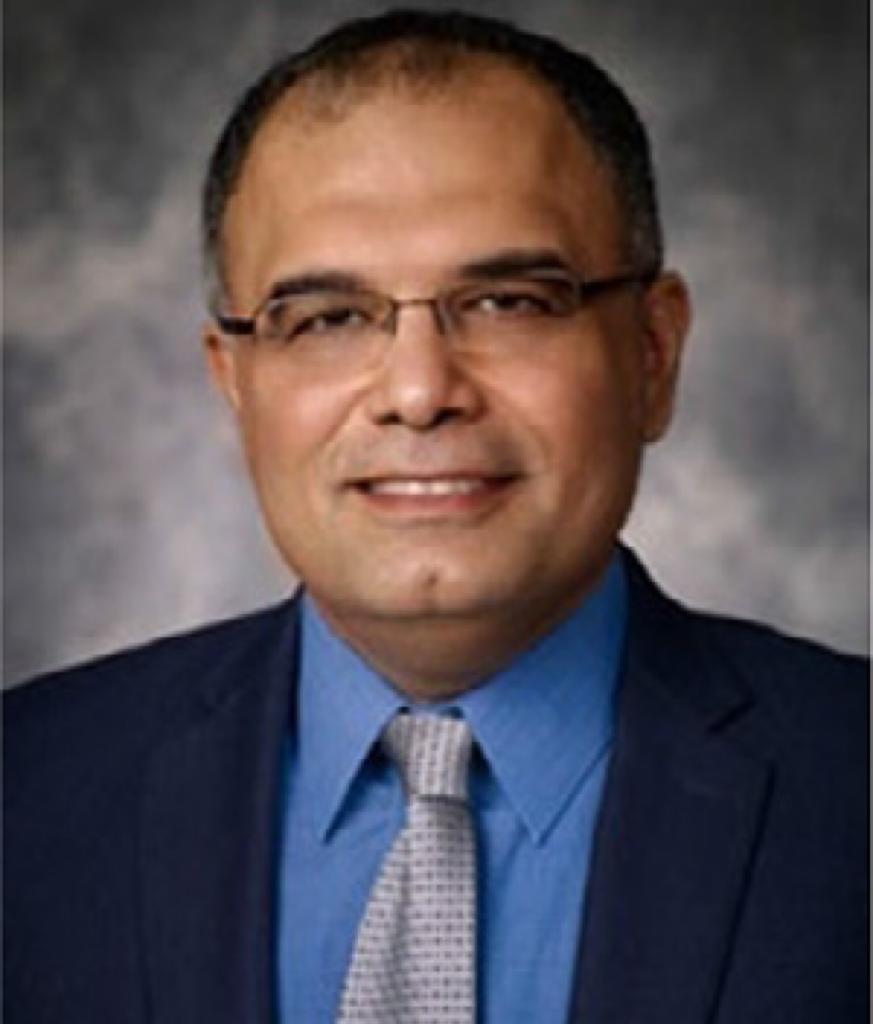}}]
{S. O. Reza Moheimani} is a professor and holds the James Von Ehr Distinguished Chair in Science and Technology in the Department of Systems Engineering at the University of Texas at Dallas with appointments in Electrical and Computer Engineering and Mechanical Engineering Departments. He is the founding Director of UTD Center for Atomically Precise Fabrication of Solid-State Quantum Devices and founder and Director of laboratory for Dynamics and Control of Nanosystems. He is a past Editor-in-Chief of Mechatronics (2016-2021), and a past associate editor of IEEE Transactions on Control Systems Technology, IEEE Transactions on Mechatronics and Control Engineering Practice. He received the Industrial achievement Award (IFAC, 2023), Nyquist Lecturer Award (ASME DSCD, 2022), Charles Stark Draper Innovative Practice Award (ASME DSCD, 2020), Nathaniel B. Nichols Medal (IFAC, 2014), IEEE Control Systems Technology Award (IEEE CSS, 2009) and IEEE Transactions on Control Systems Technology Outstanding Paper Award (IEEE CSS, 2007 and 2018). He is a Fellow of IEEE, IFAC, ASME, and Institute of Physics (UK).
Moheimani received the Ph.D. degree in Electrical Engineering from University of New South Wales, Australia in 1996. His current research interests include applications of control and estimation in high-precision mechatronic systems, high-speed scanning probe microscopy and atomically precise manufacturing. He is leading a multidisciplinary effort to develop new tools and methods for fabrication of solid-state quantum devices with atomic precision based on ultra-high vacuum scanning tunneling microscope.
\end{IEEEbiography}

\end{document}